\documentclass[10pt,letterpaper,leqno,oneside]{article}

\title{System Description for a Scalable, Fault-Tolerant, Distributed Garbage
Collector}
\author{N. Allen and T. Terriberry \\
Department of Computer Science \\
Virginia Polytechnic Institute and State University \\
Blacksburg, Virginia 24061}
\date{June 17, 2002}

\newcommand{\defstyle}[1]{\textbf{#1}}
\newcommand{\memberstyle}[1]{\ensuremath{\mathit{#1}}}
\newcommand{\msgstyle}[1]{\ensuremath{\mathcal{#1}}}
\newcommand{\paramstyle}[1]{\ensuremath{\mathtt{#1}}}
\newcommand{\varstyle}[1]{\ensuremath{\mathit{#1}}}

\newcommand{\ackmsg}[2]{\msg{{#1}\_\mbox{\tiny\,}\memberstyle{ACK}}{#2}}
\newcommand{\ackxmsg}[1]{\xmsg{{#1}\_\mbox{\tiny\,}\memberstyle{ACK}}}
\newcommand{\define}[2]{\par\noindent\defstyle{#1}- {#2}}
\newcommand{\member}[1]{\memberstyle{#1}}
\newcommand{\memberarr}[3]{\membervar{#1}{#2}[\paramstyle{#3}]}
\newcommand{\membersub}[2]{\ensuremath{{\mbox{\memberstyle{#1}\,}}_{\!\!\varstyle{{#2}}}}}
\newcommand{\membervar}[2]{\varstyle{#1}.\member{#2}}
\newcommand{\msg}[2]{\xmsg{#1}(\paramstyle{#2})}
\newcommand{\state}[1]{\varstyle{#1}}
\newcommand{\var}[1]{\varstyle{#1}}
\newcommand{\xmsg}[1]{\msgstyle{#1}}

\newcommand{\accepts}{\member{accepts}}
\newcommand{\activ}{\member{active}}
\newcommand{\activvar}[1]{\membervar{#1}{\activ}}
\newcommand{\backl}{\member{backlist}}
\newcommand{\backlvar}[1]{\membervar{#1}{\backl}}
\newcommand{\depen}{\member{dependents}}
\newcommand{\depenvar}[1]{\membervar{#1}{\depen}}
\newcommand{\done}{\member{done}}
\newcommand{\donevar}[1]{\membervar{#1}{\done}}
\newcommand{\ei}[1]{\membersub{Ei}{#1}}
\newcommand{\entryc}{\member{entrycolors}}
\newcommand{\entrycvar}[1]{\membervar{#1}{\entryc}}
\newcommand{\entryl}{\member{entrylist}}

\newcommand{\ex}[1]{\membersub{Ex}{#1}}
\newcommand{\exitl}{\member{exitlist}}
\newcommand{\exitlvar}[1]{\membervar{#1}{\exitl}}
\newcommand{\fset}{\member{fault\_\mbox{\tiny\,}set}}
\newcommand{\fsetvar}[1]{\membervar{#1}{\fset}}
\newcommand{\group}[1]{\ensuremath{\mathcal{#1}}}
\newcommand{\groupl}{\member{group}}
\newcommand{\groupvar}[1]{\membervar{#1}{\groupl}}
\newcommand{\joins}{\member{joins}}
\newcommand{\leader}[1]{\ensuremath{\overline{\pt{#1}}}}
\newcommand{\leadervar}[1]{\membervar{#1}{\member{leader}}}
\newcommand{\machine}[1]{\membersub{M}{\ensuremath{\!}{#1}}}
\newcommand{\marks}{\member{\member{marks}}}
\newcommand{\marksvar}[1]{\membervar{#1}{\marks}}
\newcommand{\mergel}{\member{merge\_\mbox{\tiny\,}list}}
\newcommand{\mergelvar}[1]{\membervar{#1}{\mergel}}
\newcommand{\msgAC}[1]{\msg{AC}{#1}}
\newcommand{\msgAR}[1]{\msg{AR}{#1}}
\newcommand{\msgARACK}[1]{\ackmsg{AR}{#1}}
\newcommand{\msgEC}[1]{\msg{EC}{#1}}

\newcommand{\msgJN}[1]{\msg{JN}{#1}}
\newcommand{\msgJR}[1]{\msg{JR}{#1}}
\newcommand{\msgMI}[1]{\msg{MI}{#1}}
\newcommand{\msgMIACK}[1]{\ackmsg{MI}{#1}}
\newcommand{\msgMR}[1]{\msg{MR}{#1}}
\newcommand{\msgMRACK}[1]{\ackmsg{MR}{#1}}
\newcommand{\msgRG}[2]{\msg{RG}{{#1},\,{#2}}}
\newcommand{\neigh}{\member{neighbors}}
\newcommand{\neigha}[2]{\memberarr{#1}{\neigh}{#2}}
\newcommand{\next}{\member{next}}
\newcommand{\nextvar}[1]{\membervar{#1}{\next}}
\newcommand{\oldpr}{\member{oldprecursors}}
\newcommand{\oldprvar}[1]{\membervar{#1}{\oldpr}}
\newcommand{\parent}{\member{parent}}
\newcommand{\precu}{\member{precursors}}
\newcommand{\precuvar}[1]{\membervar{#1}{\precu}}
\newcommand{\pt}[1]{\membersub{PT}{\ensuremath{\!}{#1}}}
\newcommand{\respo}{\member{responsibles}}
\newcommand{\respovar}[1]{\membervar{#1}{\respo}}
\newcommand{\term}{\member{term}}
\newcommand{\termvar}[1]{\membervar{\term}{#1}}
\newcommand{\vulne}{\member{vulnerable}}
\newcommand{\vulnevar}[1]{\membervar{#1}{\vulne}}
\newcommand{\xmsgCR}{\xmsg{CR}}
\newcommand{\xmsgCRACK}{\ackxmsg{CR}}
\newcommand{\xmsgCM}{\xmsg{CM}}
\newcommand{\xmsgCMACK}{\ackxmsg{CM}}
\newcommand{\xmsgEC}{\xmsg{EC}}
\newcommand{\xmsgECACK}{\ackxmsg{EC}}
\newcommand{\xmsgGI}{\xmsg{GI}}
\newcommand{\xmsgGIACK}{\ackxmsg{GI}}
\newcommand{\xmsgJN}{\xmsg{JN}}
\newcommand{\xmsgMC}{\xmsg{MC}}

\begin{document}

\maketitle

\begin{abstract}
We describe an efficient and fault-tolerant algorithm for distributed cyclic garbage collection.
The algorithm imposes few requirements on the local machines and allows for flexibility in the choice of local collector and distributed acyclic garbage collector to use with it.
We have emphasized reducing the number and size of network messages without sacrificing the promptness of collection throughout the algorithm.
Our proposed collector is a variant of back tracing to avoid extensive synchronization between machines.
We have added an explicit forward tracing stage to the standard back tracing stage and designed a tuned heuristic to reduce the total amount of work done by the collector.
Of particular note is the development of fault-tolerant cooperation between traces and a heuristic that aggressively reduces the set of suspect objects.
\end{abstract}

\section{Background}

\subsection{Need for distributed garbage collection}

Garbage collection has become a common feature in modern programming languages \cite{Wei90,CDG88}.
Interest in garbage collection is primarily due to the necessity and difficulty of proper memory management.

Memory management is necessary.
Memory is an example of a {\em non-scalable} resource, a resource with a super-linear cost function, due to the increasing costs of adding memory once a certain threshold is reached.
Systems that are persistent or are expected to be long-lived must have near-perfect resource management to maximize the quantity of non-scalable resources available.
Over a long period of time, a small, continuous loss of a non-scalable resource eventually exhausts the availability of that resource.

Memory management is difficult.
In examining memory management for a system we see that: errors are easy to create, errors are difficult to detect, and errors are very difficult to correct.
Symptoms of these errors often remain unobserved until execution reaches a point far removed from the original error or fail to show up unless the program is stressed in an unusual way \cite{Wil92}.
Tracking these errors down consumes a considerable portion of development time \cite{Rov85}.
Having conflicts in programming styles, concurrent access to the heap, and programming languages with non-deterministic execution orders further exacerbates the problem \cite{PS95,Jon96}.
Memory management also defies modularity since a module must introduce non-local bookkeeping to ensure that a particular object that the module controls is no longer needed by some other module \cite{Wil92}.

There is a requirement when using explicit memory management that allocations and deallocations be paired.
Thus, two kinds of errors can occur: deallocating memory that should not be reclaimed and failing to deallocate memory that should be reclaimed.

Of the two kinds of errors, premature deallocation is more serious.
Deallocating an object that is still in use is easy when the deallocation process lacks error checking.
Detection of this error is possible by examining other pointers to see if the memory location that they reference has been invalidated.
However, examining all other pointers in the system at each deallocation is slow and unlikely to be implemented.
Correction of a premature deallocation is not possible for most systems.

Failing to deallocate does not generally cause an immediate failure of the system but can negatively impact long-term performance.
Contributors to the failure to correctly deallocate are the ability to reassign pointers and the loss of a pointer due to scoping.
These pitfalls make indefinitely delaying the deallocation process for a segment of memory easy.
Again, detection of this error is not possible without tracing through all of the pointers in the system- a generally infeasible task.
Should the error be detected, correction of the error is easy.

Using garbage collection in a distributed environment is natural because the difficulties of explicit memory management increase as the amount of global control and information available to the memory manager decreases.
However, providing an automated but unobtrusive solution to this problem is challenging enough that garbage collection has been actively researched for over forty years \cite{McC60,Col60}.
There is a large collection of existing algorithms for the distributed case.
All of them make tradeoffs between various design goals \cite{Rod98}.
\subsection{Design goals}

The following are the requirements for designing our distributed collector.
We have listed the requirements in priority order.

\begin{enumerate}
\item
A garbage collector is said to be {\em correct} if live objects are never reclaimed.
Throughout this paper, we only consider correct garbage collectors.
\item
A garbage collector is said to be {\em complete} if garbage objects are always eventually reclaimed.
We prefer complete garbage collectors because they allow us to maximize the availability of non-scalable resources.
\item
A garbage collector is said to be {\em fault-tolerant} if the properties of completeness and correctness are preserved despite system crashes or network failures.
We prefer fault-tolerant garbage collectors because requiring failure-free operation in a distributed environment limits the scalability of the system.
\item
We prefer that the garbage collector should be efficient.
A metric for efficiency is defined below.
\end{enumerate}

Efficiency has a large number of meanings depending on the context.
We have defined, for the purposes of distributed garbage collection, an efficient system as one that:

\begin{itemize}
\item
makes little use of non-scalable resources such as network bandwidth, CPU time, memory or other storage space, and other local resources.
\item
makes little use of synchronization across machines, with the user processes, called {\em mutators}, and with the local collector.
Synchronization across machines is reduced at all costs before synchronization on the local machine.
\item
makes few delays in reclaiming garbage rather then unnecessarily deferring or unduly delaying collection.
\end{itemize}

\subsection{Basic idea}

After stating our design goals, we find that distributed garbage collectors based upon back tracing are the closest to meeting all of the requirements.
This motivates the development of a new distributed garbage collector that uses the back tracing model as its fundamental collection operation.
The problems of purely local collection and distributed acyclic garbage collection are well understood.
We therefore focus our attention on collecting distributed cyclic garbage.
Back tracing is already complete and correct so our new algorithm is built around increasing the efficiency of the collection process.
To provide fault-tolerance we must guarantee completeness and correctness despite system or network faults.
However, our efficiency concerns are centered around the fault-free case; we believe this to be the typical case.

As with other back tracing collectors, our algorithm relies on a heuristic to select candidate objects for tracing.
We design a heuristic using all of the available local state information to aggressively reduce the number of candidate objects without sacrificing promptness.
The heuristic is tightly coupled to the tracer; this allows us to avoid using network resources during the heuristic process.
A drawback to this approach is that the complexity of our heuristic is significantly greater than those in other back tracing schemes.

Once the heuristic has identified the candidate objects, we trace using a three-stage process.
The first stage is a forward trace designed to identify a subgraph reachable from the candidate objects.
Although this seems like more work then doing simple back tracing, we show that we can use the computed subgraph to significantly reduce the work done during the back tracing stage.
Additionally, this forward trace is already being used indirectly in approaches such as those based upon the distance heuristic.
The second stage is a backwards trace along the candidate subgraph to identify objects in the subgraph that are actually garbage.
The final stage performs the collection of garbage objects and cleans up the state information of objects that were found to be live.

\subsection{Object model}
We define a number of useful terms before introducing some existing distributed garbage collectors.
A distributed system is divided into one or more separate segments called {\em machines} or {\em processes}.
Each machine has its own address space for which that machine is the {\em host} and contains pieces of data called {\em objects}.
Each object may contain {\em references} to other objects or be the target of references.
Each machine contains independent processes, called {\em mutators}, that are able to modify the object reference graph.
These processes represent the actions taken by the user's application.

References to objects on the same machine are called {\em local references}; references to objects on a different machine are called {\em remote references}.
Objects are {\em distributed objects} if they are the target of at least one remote reference.
Otherwise, the object is considered to be a {\em local object}.
Special objects, identified as {\em roots}, are never considered to be garbage.
Examples of roots are global variables and the runtime stack for a thread of execution.
A {\em local root} is a root on the same machine as a particular object.
An object is {\em locally reachable} from a set of objects if that object is reachable by traversing local references from some object in the set.
If no set is specified, the set of local roots is assumed.

Remote references are represented by a pair of distinguished objects called an {\em entry item} and an {\em exit item}.
A distributed object, \var{x}, has on its host a single entry item, \ei{x}, for that object.
\ei{x} has a local reference to \var{x}.
A machine containing a remote reference to \var{x} has an exit item, \ex{x}.
\ex{x} contains a {\em locator} for \var{x} which is used to identify \ei{x}.

There are four basic operations for remote references.
First, a reference may be created by the machine that hosts an object by sending a locator to a remote machine.
Next, a reference may be duplicated if one machine sends a locator to another machine.
This requires interaction only with a machine that already has a locator for that object.
A reference may be traversed by passing its locator as a parameter of a remote procedure call on the host machine.
This creates a new local root for that object in the thread that handles the call.
Finally, a reference may be destroyed by deleting the corresponding exit item.

\subsection{Existing distributed garbage collectors}

The majority of distributed garbage collectors are based upon reference tracking collectors, tracing collectors, or hybrid collectors.
Reference tracking collectors maintain counts or lists of references on the local system to determine if an object is garbage.
Tracing collectors follow the references from the root sets of all of the local systems to determine if an object is garbage.
Hybrid collectors use both reference tracking and tracing information to determine if an object is garbage.

Each kind of distributed garbage collector is further refined by the degree of {\em locality} of the particular algorithm.
An algorithm has a high degree of locality if collecting a garbage cycle requires cooperation of only the machines in the garbage cycle \cite{Rod98}.
Utilization of other machines during the collection process does not reduce the degree of locality as long as the other machines can fail at any time without preventing collection.
This is contrary to other definitions of locality \cite{ML97} which disallow the involvement of any machines not hosting an object in the garbage cycle.
However, the algorithm presented therein uses a non-local heuristic which would violate the given claim of locality using this stricter definition.

\subsection{Reference counting}\label{sec:reftrack}
Reference tracking and other similar variants are the most common forms of direct identification of garbage.
Direct identification of garbage detects objects which are not reachable from a live object.
These detected objects are marked for reclamation.
The objects which are not marked must then be live.
Reference tracking is attractive in distributed systems for two reasons.
The operation of the collector is interleaved with that of the mutators and does not require any global information.
Thus, reference tracking has good concurrency and a high degree of locality.

Reference tracking is usually considered to be too expensive for uniprocessor systems because the cost is proportional to the total amount of work done by the system.
In distributed systems however, that overhead is applied only when distributed references are created or copied.
These operations already involve a communication overhead to send the references to different machines and tend to be much rarer.

\subsubsection{Basic reference counting}
The idea behind reference counting is relatively simple \cite{Col60}.
Each object keeps a count of the number of references to it.
Creating a new reference increments the count; destroying a reference decrements the count.
When the count reaches zero, the object is garbage and is reclaimed.

Deleting a reference can cause an unbounded delay when a large portion of the reference graph is reclaimed.
A solution is to place reclaimed objects in a queue instead of immediately decrementing the count of objects to which they refer \cite{Wei63}.
When the storage space for an object is about to be reused, the old references from the object are cleared, and the counts of the referents are decremented.
This limits the number of counts decremented in a single deallocation to the number of references stored in an object.
Therefore, this modification provides better mutator concurrency.

When using this algorithm in a distributed system, only remote references need to be tracked.
All of the references on the same machine to one remote object can share a single exit item.
No additional overhead is imposed on local operations.
Each machine runs its own local garbage collector using any of the uniprocessor techniques to handle local references.
The local collector considers objects with a positive number of remote references to be live.
When the remote reference count of an object reaches zero, the object may be reclaimed as soon as the local collector determines that there are no local references to the object.

Whenever a remote reference is duplicated or deleted, an increment or decrement message is sent to the host of the object.
However, messages may be lost, delayed, or reordered.
If an increment message is sent followed by a decrement message, then the object may be reclaimed unsafely if the decrement message arrived prior to the increment message.
Additionally, if a decrement message is lost or an increment message is duplicated, then the object to which the message referred can never be reclaimed.
If a process crashes, then decrement messages are never sent for any contained remote references which also prevents those objects from ever being reclaimed.
Finally, reference counting is not complete even without any failures.
If two objects hold references to each other, neither object can ever be reclaimed.

\subsubsection{Acknowledgement messages}
Depending on which machine sends the increment message when a reference is copied from one machine to another, there are two types of race conditions between increment and decrement messages.
Both race conditions are avoided by using acknowledgement messages \cite{LM86}.

If the receiver of a reference is responsible for sending the increment message, a decrement-increment race condition occurs when a machine sends a remote reference to another machine and then deletes its own copy of the reference.
If the sending machine had the last reference to that object and the decrement message arrived before the receiving machine's increment message, the object is reclaimed unsafely.
To avoid this, when a machine receives a copy of a remote reference, an acknowledgement request is sent along with the increment message.
The host of the object then sends an acknowledgement to the machine that is sending a copy of the reference.
Any decrement messages must be queued until this acknowledgement is received.

If the sender of a reference is responsible for sending the increment message, an increment-decrement race condition occurs when a machine sends a remote reference to another machine that then deletes the reference.
If the sending machine had the last reference to the object and the receiving machine's decrement message arrives before the sending machine's increment message, the object is reclaimed unsafely.
To avoid this, when a machine wants to send a reference to another machine, an acknowledgement request is sent along with the increment message.
The machine then waits until the host responds with an acknowledgement.
Once the sending machine has received this acknowledgement, a copy of the reference is sent to the receiving machine.

Even with these acknowledgement messages, reference counting is still not safe from lost or duplicated messages.
Other reference tracking schemes are more resilient to these kinds of message failures.

\subsubsection{Weighted reference counting}
Weighted reference counting is a variant of reference counting that eliminates the increment message.
This improves message efficiency and avoids race conditions \cite{Bev87,WW87}.
A weight is associated with each reference in addition to a count.
The system keeps the sum of the weights associated with all of the references to an object as an invariant.
This sum is also known as the object's reference count.

When an object is created, its reference count is initialized to some maximum value.
The first reference to the object receives that value as its weight.
When a machine wants to copy a reference, the weight of that reference is divided evenly with the new reference.
When a reference is deleted, the weight is sent in the decrement message.
The weight is then subtracted from the object's reference count.
When an object's reference count reaches zero, no references to the object can exist, and the object is reclaimed.

Although this system reduces the number of messages sent and eliminates race conditions, extra space is required to store the weight with each reference.
In practice, only the logarithm of the weight need be stored.
However, the weight must be expanded to perform the subtraction when a reference is deleted.
Also, the algorithm is still not resilient to message loss, message duplication, or process failure.

The most serious drawback is that once a weight reaches some minimum value, the weight cannot be further subdivided.
One solution is to request additional weight from the owner of the object.
However, this is equivalent to using increment messages.
Another solution is to create an indirection object.
An indirection object has its own reference count and forwards messages to the original object.
In the worst case, a long chain of indirection objects are required which greatly increases the number of messages required for a single computation.

By weakening the invariant so that an object's reference count is greater than or equal to the sum of the weights of all references to the object, both message loss and the problem with reference duplication are corrected \cite{Dic92}.
When a reference's weight can no longer be divided, the weight is replaced with a special null weight value.
The null weight value can be copied any number of times.
An object's reference count may exceed the sum of the weights of its references which prevents collection.
However, reference counting is not complete, and using null weight values does not hinder a scheme where reference counting is augmented by some other technique.
Message duplication might still cause an object to be reclaimed unsafely.

\subsubsection{Indirect reference counting}
Indirect reference counting avoids indirection objects but requires additional memory.
Instead of associating a weight with each reference, a reference gets its own reference count.
When a reference is copied, the local reference count is incremented and the copy of the reference stores a locator for the reference from which it was copied.
A reference cannot be deleted until its local reference count reaches zero.
When the local reference count reaches zero, a decrement message is sent to the reference from which the copy was made.

Increments are done locally so that there are no race conditions associated with indirect reference counting.
The process however, is still not resilient to message loss, message duplication, or process failure.
The same technique can also be used to reduce the message overhead of more reliable forms of reference tracking such as the one presented in the next section.
The smaller number of messages sent is offset by having larger messages.
Each reference now contains a weak pointer to the actual object and a strong pointer to the reference from which this reference was copied.
In addition to the space overhead required to store both pointers, each machine may also be required to keep unreachable references because copies of those references are still needed by other machines.

\subsubsection{Reference listing}
Reference listing handles lost messages, duplicated messages, and process failure by replacing an object's reference counter with a list of processes that hold a reference to the object.
This change adds a significant space overhead to maintain the list but provides a robustness lacking in other reference tracking schemes.
Instead of sending increment and decrement messages, insert and delete messages that contain the identity of the process holding the reference are sent when a reference is copied or deleted.

Lost messages are replaced by resending failed insert or delete messages.
Alternatively, the entire list of references one machine has can periodically be sent to another machine.
Although distributing the entire list of references is simple to implement and does not require sending acknowledgement messages, individual messages can be very large whereas insert, delete, and acknowledgement messages are typically small.
Furthermore, in a networking environment where large packets must be broken up to be transmitted and delivery of any one packet is unreliable, a large list of references is unlikely to be transmitted without error.

Duplicate messages are safely ignored since a process can only be added or deleted from a reference list once.
Race conditions are handled by using acknowledgement messages or a scheme such as indirect reference counting that avoids sending insert messages.
Process failure is handled by removing from the list all of the references originating from processes which are known to have failed.

Reference listing introduces an additional race condition.
If a machine deletes a reference to an object and later reacquires a reference, the delete message might not arrive until after the insert message has arrived.
In this case, the insert message appears to be a duplicate and is discarded.
When the delete message is received, the object is reclaimed unsafely.
This problem is avoided by timestamping insert and delete messages and storing in the reference list the largest timestamp from an insert message.
If a delete message is received with a timestamp that is smaller than the timestamp in the reference list, the delete message is ignored.

Reference listing is resilient to process failures as well as message loss, duplication, and reordering.
The cost of this safety is additional space overhead for maintaining the reference lists.
This is probably acceptable if used only for objects to which there are remote references.
As with all reference tracking techniques, reference listing has good scalability, concurrency, and promptness.
A major drawback is that reference listing is not complete since cyclic garbage cannot be collected.

\subsection{Tracing}\label{sec:tracing}
Tracing algorithms are the most common kind of collectors based upon indirect identification of garbage.
Tracing collectors start at roots and mark live objects that are encountered by traversing references.
All unmarked objects must then be garbage.
Tracing algorithms often require much synchronization with remote machines and with the local mutators.
Furthermore, the trace grossly violates the property of locality by passing through all live objects.
However, unlike reference tracking schemes, tracing schemes can collect distributed garbage cycles.
Tracing algorithms are among the most successful for uniprocessor machines; having a poor degree of locality, and increased complexity, makes an extension to distributed environments less appealing.

\subsubsection{Basic global tracing}
Tracing algorithms typically operate by using two distinct phases \cite{Rod98}.
The first phase, mark, traverses all objects reachable from a root.
The second phase, sweep, reclaims all objects that were not marked by the first phase.
These reclaimed objects must be garbage because they cannot be reached from any root.

A machine that is running low on storage space initiates garbage collection by sending a request to a controlling machine called a coordinator.
The coordinator ensures that all of the machines suspend their mutators so that the reference graph cannot be modified during the first phase.
Then, each machine marks all of its live objects and sends a mark request message for each encountered remote reference.
Machines alternate between an active state, where they are marking objects, and an idle state.
An active machine sends mark messages until no more live objects are encountered and then transitions to the idle state.
Receipt of a mark message by an idle machine makes the machine active.
The mark phase is complete when all of the machines are idle.
Existing distributed termination detection algorithms can determine when all of the machines are idle.
More complicated, specialized schemes use methods that are less expensive or better deal with mutator concurrency.

After the mark phase is complete, each machine resumes the local mutators and begins the sweep phase.
The objects reclaimed in the sweep phase cannot be reachable from any live object so synchronization between the sweep phase and the mutators is not required.

This method seems unsatisfactory even as an initial approach.
The cooperation of every machine in the system is required; there is neither fault-tolerance nor scalability.
If one machine needs to reclaim garbage, then all of the other machines must stop to reclaim garbage as well.
Also, no work is done while the mark phase is in progress.
The synchronization of the mark phase requires all idle machines to wait until the last machine is finished before resuming work.
We discuss solutions for some of these problems in the following sections.

\subsubsection{Marking-tree collector}
The marking-tree collector was one of the earliest distributed tracing collectors and is based upon a concurrent version of the mark-and-sweep algorithm for uniprocessor systems \cite{HK82}.
Unlike the naive approach, the identification of garbage and mutator operations are executed concurrently.

Assume that there is a single root from which marking begins in the entire distributed reference graph.
Each machine maintains a queue of mark and mutator tasks which are executed alternately.
During a mark task, a tree is created for additional mark tasks of objects that are not already marked.
When a mark task completes, an acknowledgement is returned to the spawning task.
A mark task is complete when all of its spawned mark tasks have returned an acknowledgement.
Once the root receives acknowledgements from all of its spawned mark tasks, the mark phase is complete.
An extension to computations with more than one root is done by declaring termination when the mark task of every root is complete.
Equivalently, we construct a new tree with all of the roots present as direct children of some new root.
The trees in this scheme have a large space overhead which is proportional to the number of live objects in the system.

Objects are colored white, gray, or black.
An object is colored white if the object has not been visited in the mark phase.
At the end of the mark phase, all white objects are garbage.
An object is colored gray if the object has been visited in the mark phase but some of the object's mark tasks are not complete.
An object is colored black if the object was either allocated during the current mark phase or visited in the mark phase and all of the object's mark tasks are complete.
At the end of the mark phase, all live objects are colored black.

Mutator and mark tasks compete to modify objects by locking all of the objects that they intend to modify.
Mutator operations must preserve the invariants associated with the color of an object.
For example, copying a reference for a white object into a black object requires spawning a mark task for the white object and waiting for acknowledgements.
This can introduce significant delays if a large portion of the reference graph is traversed.
The mark phase must be synchronized across every machine, but we still have some concurrency between the mutator and the collector.

\subsubsection{The Emerald system}
The Emerald system is also based upon a three color tracer but provides better mutator concurrency and achieves completeness despite encountering machines that are temporarily unavailable.
The key difference compared to the marking tree collector is that objects may be colored black if, instead of waiting for the entire subtree of mark tasks to complete, all of the objects to which they refer are colored gray.
Mutator concurrency is supported by an object fault mechanism similar to page faults in a virtual memory system.
Copying a white reference into a black object is not explicitly prevented but does color the white object gray.
A read barrier that protects all gray objects is then imposed.
Traversing a reference to a gray object invokes a fault handler in the collector that then colors gray all of the objects reachable from the gray object.
The object that the mutator was trying to reach may then be colored black.
Using this system, large portions of the reference graph are never traversed to satisfy a single reference of a gray object.

A black object cannot have a reference to a white object, and all gray objects are colored black before allowing a mutator to read them.
Thus, a mutator which operates on black objects cannot obtain a reference to a white object.
Each machine maintains a non-resident gray set to avoid arbitrary communication delays while waiting for a remote object to be colored gray.
This set contains remote references for which mark messages have been sent but not acknowledged.
These references are treated as references to gray objects, but the coloring of the actual objects may be postponed.
After the remote machine colors the object gray, the mark message is acknowledged, and the reference is removed from the non-resident gray set.
Finally, objects that are remotely invoked are immediately colored gray to prevent a black object from accessing a white remote object.

The diffusing tree algorithm from the previous section cannot detect the termination of the mark phase.
Termination is instead detected using a two phase commit protocol that is resilient to process and message failures but requires that each process know of every other process in the system and decreases scalability.

\subsubsection{Tracing with timestamps}
Timestamps are also, in a similar fashion to colors, used in tracing \cite{Hug85}.
A timestamp, instead of a color, is assigned to each object in the marking phase.
The marking phase runs perpetually increasing the timestamps of live objects.
The timestamp of a garbage object eventually stops increasing.
A global threshold is computed from the clocks of each machine, and objects with a timestamp below that threshold are reclaimed safely.

A local collector that is also a tracing collector can perform the actual marking operations.
Each machine keeps a clock such as Lamport's logical clock \cite{Lam78}.
The start time of a local collection, GC-time, is recorded.
Local roots and objects without a timestamp are assigned the GC-time.
Then, the local collector propagates timestamps so that an object is marked with the maximum of the timestamps of the objects from which that object is reachable.
This is done by processing the mark requests in a descending order.
Timestamps are forwarded across remote references in timestamping messages.
The local collector maintains mutator concurrency so that standard uniprocessor techniques can be used.

Each machine maintains a threshold value called a redo timestamp.
When a machine has completed a local collection, has not received any timestamping messages, and has received acknowledgement messages for all of the timestamping messages that were sent out, the redo timestamp is set to the value of the GC-time.
The minimum of all of the redo timestamps is the global threshold.
The propagation of timestamps acts as mark phases operating in parallel.
Mark phases corresponding to a timestamp below the global threshold are complete.

Computing the global threshold requires a costly termination detection algorithm and involves the cooperation of every machine.
Thus, this algorithm has no locality.
If a machine fails, then the global threshold never increases beyond the last reported timestamp of that process.
Therefore, collection eventually stops.
If a machine is slow or rarely triggers a local collection, then the threshold may not increase far enough for the other machines to reclaim needed space.
This is true even if the slow machine has no remote references.

\subsubsection{Logically centralized tracing}
Another approach is to offload the tracing of the reference graph to some centralized service \cite{LL86}.
When a machine performs a local collection, the centralized service is notified about the remote references that are encountered.
Each machine periodically queries the centralized service about objects that are not locally reachable.
If an object is not locally reachable and is reported to have no remote references, then the object is reclaimed safely.

As before, garbage cycles that span machines are collected using timestamps.
A centralized service maintains a copy of the timestamps and computes the threshold time to avoid using a costly termination detection algorithm.
Unlike the previous section, the timestamps are used only to collect cyclic garbage.
The centralized service can be replicated to achieve fault-tolerance, but doing so does not increase scalability.
At least one of the services still must be able to communicate with all of the processes to collect garbage.
Furthermore, the space required is proportional to the total number of remotely reachable objects in the entire system.

\subsection{Hybrid schemes}\label{sec:hybrid}
Reference tracking systems have a good degree of locality, can be made fault-tolerant, are prompt, and are highly concurrent.
The principle drawback is that reference tracking cannot collect cyclic garbage.
Tracing algorithms are complete but neither scalable nor prompt since an entire trace must be completed before any garbage is collected.

Several systems trade completeness with efficiency and fault-tolerance by assuming that distributed cycles are uncommon and can be ignored.
This assumption is true for short-lived systems that have sufficient memory to allow for storage leaks.
In long-lived distributed systems however, even a small memory leak accumulates until the resource is exhausted.
Furthermore, cycles are commonly created in systems using client-server communication, replication, or mobile computing.
Without a cyclic distributed garbage collector, cycles must be broken manually, but we know of no good methodology for doing so \cite{Rod98}.

A solution is to use reference counting for collecting acyclic garbage and a tracing algorithm for collecting cyclic garbage.
However, on a smaller scale, this approach still has the same drawbacks as using only a tracing collector.

\subsubsection{Tracing in groups}
Tracing in groups uses groups of processes to reduce the problems of scalability and fault-tolerance associated with standard tracing algorithms \cite{LQP92}.
Reference counting is used to collect the acyclic garbage; a mark-and-sweep collector is used to collect the garbage cycles within a group.
Groups do not need to be fixed at process creation.
Instead, a group is configured dynamically and can share machines with other groups.
Collections in two different groups can be run on the same machine in parallel although this requires additional computation by the local collector.

The degree of locality of the algorithm depends on how the groups are formed.
A garbage cycle which does not span groups requires only the cooperation of machines in the group.
The algorithm is fault-tolerant because when a machine fails, the remaining machines form a new group and restart the trace.
We also improve promptness because a small group containing a garbage cycle can collect that cycle without waiting for a trace to complete on all of the machines in the system.
A problem with this algorithm is that there is no clear method for forming groups.
A tree-like hierarchy of groups is suggested to ensure that every garbage cycle is covered by some group.
However, the smallest group containing a particular garbage cycle may contain many more machines than are involved in the cycle.

\subsubsection{Local tracing}
Local tracing algorithms combine weighted reference counting with a tracing algorithm \cite{LJ91}.
Unlike the previous tracing algorithms, tracing begins from objects that are suspected to be garbage instead of the roots of the reference graph.
Suspect objects are identified by using a heuristic.
The proposed heuristic traces the referents of deleted references.
This form of tracing is similar to an earlier technique called trial deletion \cite{Ves87}.

The tracer decrements the reference counters of the objects that are reachable from the suspect object.
The suspect object is garbage if, at the end of the trace, its counter has been reduced to zero.
Otherwise, the values of the decremented reference counters are restored.
This algorithm does not have perfect locality since some of the traced objects may be live.
However, only objects reachable from objects whose counts were modified are visited.
Therefore, this algorithm is more scalable than global tracing.

The first algorithm of this kind also had the drawback that the local collector was required to be based upon reference counting; a later proposal eliminated this requirement \cite{MKITHN95}.
We must also ensure that two different traces do not simultaneously modify the reference count of an object.
This requires either a form of global synchronization or extra storage for each trace to keep a separate count.
In practice, both solutions are likely to be expensive.

\subsubsection{Object migration}
Object migration collects garbage cycles by moving all of the objects on a cycle to one machine.
Then, a local collector detects the garbage cycle and reclaims all of the objects.
The collector first suspects objects of being garbage by using a heuristic.
A suspect object is migrated to a machine containing a reference to that object.

This algorithm has locality because only the machines that originally contained an object on a garbage cycle are required to migrate objects on that cycle.
However, there are serious drawbacks to this approach.
Indirection objects are created so that machines can refer to a migrated object without updating references.
Eliminating these indirection objects requires additional message passing or the construction of unique identifiers that track objects.
Also, migration is an expensive operation that requires the transmission of a large amount of data which will eventually be discarded as garbage.
Migrating objects may exhaust the resources of a single machine.
All of the objects on a garbage cycle must be contained on one host prior to any garbage being collected.
Finally, object migration may not be supported on a heterogeneous network where hosts have different representation models for objects.

\subsubsection{Train collection}
The train algorithm is an adaptation of a complete, concurrent and scalable uniprocessor algorithm \cite{MMH96,HMMM97}.
The algorithm partitions the distributed address space into disjoint regions, called cars, which reside on a single machine.
A local collector reclaims only garbage cycles within a car, but cars are grouped into trains to collect garbage that spans between cars.
Garbage collection for a train runs by moving reachable objects to other trains.
Eventually, a train contains nothing but garbage, and all of the objects in the train are reclaimed safely.

Reference listing tracks cars that contain a reference to an object.
Cars that have a common reference to an object join together into a train.
The first car that joins a train is designated the master and records when cars join or leave the train.
The entire train is collected when there are no references into the train from cars outside of the train.
We detect this using a distributed termination detection protocol that requires only machines in the train.

Only cars in a train are required to collect garbage so train collection has a high degree of locality.
This is not optimal because a single train can contain more than one garbage cycle.
The algorithm is made fault-tolerant by reforming a train when a car in the train fails.
Only the reference listing operation needs to be synchronized with the mutator; this approach provides good mutator concurrency.
A drawback is that moving an object to a new car or train requires updating all of the old references to the object.

\subsubsection{Timestamp packet distribution}
A different approach to distributed garbage collection is to construct causality inferences between mutator events that modify the reference graph \cite{LC97}.
A history of causality events allows some kinds of garbage cycles to be collected without determining a global state.
This algorithm is a direct method of identifying garbage with the propagation of the causal history of an object similar to tracing with timestamps.

The causal relations between events are represented using vector times.
These relations maintain a machine's view of every process in the system when an event occurs \cite{Mat89}.
Causality information in the form of clock time is transmitted between machines when a message is sent.
An underlying reference listing scheme propagates this information until an object is determined to no longer be reachable from a root.

This algorithm has the benefits of reference tracking, is complete, and does not have the drawbacks of tracing.
Time information is required only to pass through the objects in a garbage cycle, so timestamp packet distribution has locality.
However, each object must maintain a list of vector times corresponding to the paths from which the object is reachable.
Each vector time requires space proportional to the total number of machines in the distributed system.
Thus, the storage space overhead of the algorithm is very large.
Additionally, determining the inferences used by the algorithm is complicated.
The level of concurrency possible between the collector and mutator is unclear.

\subsubsection{Partial tracing}
Partial tracing resolves some of the issues with tracing in groups and local tracing \cite{Rod98}.
We use reference listing to collect acyclic garbage, identify suspect objects using a heuristic, and perform traces from suspect objects to collect garbage cycles.
However, a three phase mark-and-sweep tracer is used instead of trial deletion.

The first phase, mark, colors red all of the objects that are reachable from the suspect object but not from a local root.
The objects colored red are the suspect subgraph.
The second phase, scan, colors green all of the red objects that are reachable from objects outside of the suspect subgraph.
Objects colored green are conservatively believed to be live.
Colors are propagated from an incoming remote reference to an outgoing remote reference by a local garbage collector that is based upon tracing.
The final phase, sweep, reclaims objects that are still colored red.

One advantage of this algorithm is that different partial traces can cooperate to identify garbage.
When two traces meet in the mark red phase, they form a group of cooperating traces.
The members of the group proceed to the scan phase once every group member has completed its mark phase.
Two traces are in the same group if the intersection of the suspect subgraphs is non-empty.
A trace in the mark phase can safely ignore parts of the subgraph that are used by another trace in the scan or sweep phase.
If the suspect subgraphs of the two traces contained objects on a common garbage cycle, they would have met in the mark phase.
During the mark and sweep phases, the mutator and the tracing algorithm require no synchronization since the two processes have no contention.
During the scan phase, the mutator operates concurrently with the collector by using access barriers.

Partial tracing provides a mechanism for forming groups of machines from the structure of the reference graph.
However, the algorithm still does not have locality since the mark red phase traces through live objects.
Completeness is traded against locality by limiting the mark red phase's extent or the traces that can join a group.
Doing so increase promptness, fault-tolerance, and scalability but may fail to reclaim all of the garbage.
We know of no preferred method for finding an optimal balance between these properties.

The promptness and fault-tolerance of the algorithm depend on the heuristic that identifies suspect objects.
The heuristic may impose undue delays between the time that an object becomes unreachable and the time that the object is suspected.
If a machine belonging to a group that contains a garbage cycle fails, then the current trace for that group fails.
The heuristic determines if an object that participated in a failed trace becomes suspect; the algorithm is fault-tolerant only if the heuristic eventually re-suspects objects that are in a garbage cycle.

\subsection{Back tracing}
The only remaining hybrid algorithm that preserves locality is back tracing \cite{Fuc95,ML97,Rod95,RR97}.
Reference listing is used to collect acyclic garbage, and a heuristic identifies the suspect objects that remain.
However, instead of following the references contained in a suspect object, the tracing algorithm traverses backwards along the references to the suspect object.
If no roots are found among the objects that can reach the suspect, then all of the traced objects are reclaimed.
Faulty machines encountered in the back trace are assumed to contain live objects; this makes the algorithm safe in the presence of failures.
Message identifiers and acknowledgements protect the tracing from message duplication or loss.

Back tracing must overcome four main challenges to be viable \cite{Fuc95}:
\begin{itemize}
\item
A good heuristic is required to detect suspect objects.
This same challenge is faced by many of the hybrid algorithms and is addressed below.
\item
There is considerable overhead on the local collector in computing the backwards reference information.
Later proposals introduce an algorithm which visits each object only once \cite{ML97}.
However, the space required for storing full reachability information between incoming and outgoing references is prohibitive.
This is made tolerable by restricting traces to suspect objects, increasing the dependence on a good heuristic for identifying suspects.
Furthermore, additional messages are required to determine which incoming references come from suspect objects.
Caching of backwards reference information avoids recomputation for objects that have not changed between traces; this does not help the space overhead.
\item
Multiple back tracings must be synchronized with each other.
The current literature allows multiple tracings to exist independently but provides no means for overlapping tracings to share information and avoid repeated work \cite{ML97,Rod95,RR97}.
\cite{Fuc95} suggests imposing a total ordering on the different tracings and
blocking lower priority ones until higher priority ones have completed.
This reduces extra work in the case that the objects being traced are in fact garbage if the highest priority trace is the first to encounter an object.
If a lower priority trace arrives first, then that trace traverses part of the graph that must be re-traversed by the higher priority trace.
If an object is actually live, then it is traversed by all of the traces.
\item
The back tracing must be synchronized with the mutators.
Recent work demonstrates that this can be done without imposing any additional remote synchronization by adding a barrier to reference creation and remote method invocation.
\end{itemize}

Back tracing has a good degree of locality, fault-tolerance, and concurrence that makes it attractive as a complete solution.
However, the wasted work by repeated tracing and the space overhead might be excessive in a real application.
Another problem is that a mutator can prevent the collector from ever terminating by indefinitely creating new backwards paths while a trace is in progress.
This would destroy promptness, and garbage would accumulate until the system exhausts its storage.
Such a situation is likely to be rare and is avoided by abandoning the trace after traversing a large number of references; the work is wasted.

\subsection{Heuristics}
Many of the hybrid algorithms perform their tracing from objects that they suspect of being garbage instead of from the roots of the reference graph.
For these algorithms, a good heuristic for determining when an object is suspect is important; wrongly suspecting an object can lead to much wasted work.
The heuristic originally proposed for local tracing was to suspect the referent of a deleted reference.
This heuristic was applied to any reference and not limited to the deletion of remote references.
If only remote references were considered, then modifications to the object graph on a single machine might disconnect a cycle from the live portion of the graph without causing a remote reference to be deleted.
This heuristic is clearly far too expensive.

One property every heuristic should have is that no suspect object may be reachable from a root on the same local machine.
This is computed by the local garbage collector and does not require any message passing to determine.
An object that satisfies this condition is said to be {\em isolated}.
Since every garbage object is isolated, the condition can be added to any heuristic \cite{Rod98}.

The simplest viable heuristic is to use isolation as the only condition for suspecting objects.
However, we believe that in long-lived distributed systems there are many objects that are isolated but are not garbage.
This can lead to an unbounded number of failed traces.

To alleviate this problem, imposing a delay between traces has been suggested.
The delay is increased with each failed trace- similar to generational schemes from uniprocessor algorithms which take advantage of the tendency that objects that have survived previous collections are garbage far less often than newly allocated objects.
However, the same pattern may not hold on a long-lived distributed system where creating and using distributed objects is more expensive.
Thus, many objects might not be accessed for long periods of time.
In this case, such a delay imposed on most of the objects in the system would hurt promptness while still wasting an unlimited, albeit smaller, amount of work on failed traces.

Another technique, the {\em distance heuristic}, is stable for unchanging portions of the reference graph \cite{ML95}.
If the structure of the reference graph does not change, then the heuristic does not cause any new traces to begin.
However, the distance heuristic uses messages to determine this information, so it is more costly than previously suggested heuristics which use only information already available at each machine.

The key idea is to estimate the shortest distance of an object from any root; distance is the minimum number of remote references that must be traversed to reach that object.
A remotely reachable object maintains a distance for each remote reference to it.
Newly created remote references have a conservative estimate of one.
Roots have an associated distance of zero.
The local collector propagates distance estimates from local roots and remotely reachable objects to remote references.
Changes in the distance estimate of a remote reference are sent to the referent, and the object updates its copy of the estimate accordingly.
The key observation is that for live objects, the distance estimate eventually stabilizes at the actual distance; along garbage cycles, the distance estimate continuously increases.

This is similar to tracing with timestamps but does not suffer from some of the drawbacks.
Mainly, the failure of a single node does not prevent the distance estimates of objects in a garbage cycle that doesn't pass through that node from increasing.
This allows the system to set a suspicion threshold distance.
Objects with a distance estimate above the threshold are considered suspect.
The threshold can be modified dynamically if too many live objects are being traced or even be modified on a per-object basis.
Furthermore, multiple thresholds can provide multiple levels of suspicion.
For example, back tracing uses one threshold to allow objects to start a trace and a smaller threshold to allow the trace to pass through an object which helps alleviate the problem of multiple simultaneous back tracings on a single cycle.

There are still several problems with the distance heuristic.
One problem is that the heuristic sends extra messages for all of the remotely reachable objects even if they are not likely to be garbage.
For example, when they are locally reachable.
Another is that every cycle, no matter the size, must propagate at least as many messages as the conservatively large threshold value before becoming suspect.
This means that collecting a cycle involving only two or three objects requires the same amount of work as collecting a cycle involving hundreds or thousands of objects.
The effect is to trade one form of wasted work for another and sacrifice promptness as well.
For algorithms that only approximate locality, such as partial tracing, or which involve a significant amount of work per trace, such as object migration, this tradeoff might be acceptable.

The distance heuristic automatically sends information about suspect objects forward through the reference graph.
This is exactly what back tracing requires to limit the impact on the local collector.
The forward communication of the distance heuristic, or any algorithm which communicates when an object becomes suspect, is essentially performing a forward trace of the reference graph starting at suspect items.
The distance heuristic wastes a lot of work by sending information for non-suspect objects and by repeatedly tracing small garbage cycles until the objects' distances rise above the threshold.
Our algorithm implements this forward tracing as a distinct step and uses a heuristic that avoids much of the wasted effort of the distance heuristic.
Furthermore, independent tracings cooperate to avoid repeated work.

\section{Proposed algorithm}

Our distributed garbage collection algorithm is based upon the standard reference listing scheme.
Reference listing is easily integrated into the local collector and, at the expense of some memory overhead, is resilient to both message and process failure.
Reference listing allows us to collect all non-cyclic garbage.
The scheme is augmented with a tracing technique to collect garbage in distributed cycles.

There are three phases for detecting distributed cycles: mark gray forwards tracing, mark black backwards tracing, and sweep collection The algorithm begins tracing when a heuristic detects a suspect object, \var{x}, and creates a new partial trace \pt{x}.
The tracer colors objects to indicate the last phase that they have completed.
White objects have not yet been visited; gray objects have been visited by a trace in the first phase; black objects have been visited by a trace in the second phase.
In the third phase, an object is either garbage and is deleted or will be colored white again.
A fourth color, purple, indicates that modifications to the reference graph were made during the first two phases.

\subsection{Network model}

This section provides a brief overview of the assumptions we make about the distributed environment.
In addition to the object model mentioned earlier, our network model requires the following characteristics.

\begin{itemize}
\item
A process must be uniquely identifiable.  One source of identification may be the network address.
\item
A process may be added to the system at any time.
\item
A process may be removed only if there are no distributed references to or from that process.
\item
Any pair of processes must be able to establish a communication channel between them.
\end{itemize}

The consensus problem is unsolvable with faulty nodes under the assumption of reliable asynchronous communications \cite{FLP85}.
Distributed termination detection is harder than the consensus problem \cite{WLT92}, and distributed termination detection is reducible to garbage
collection \cite{TM93}.
Hence, to ensure reliable operation of a garbage collector, the underlying network model must be strengthened to account for node failures.

We assume that the communication network provides three additional services:
\begin{enumerate}
\item all-pair reliable end-to-end communication

Every process is able to communicate with every other process.
Messages are delivered to their destinations without errors, loss, or duplication.
If the destination process crashes, then the message is said to be undeliverable and is discarded without notifying the sender.
For our purposes, a failed process cannot send or receive any messages.
We do not handle the case of Byzantine processes that sends arbitrary messages.

\item failure detection

A process may request that notification be sent if another process fails.
Every time a process fails, the network informs any process that requested notification of the failure.
Notifications do not arrive in any deterministic order but must be delivered a finite time after the process fails.

\item fail flush

A process can request that the communication channel between itself and a process that is known to have failed be cleared.
This returns all messages sent before the failure occurred and which have not yet been delivered.
\end{enumerate}

This model separates the problem of forming a consensus about whether a process has failed, known to be impossible, from that of designing an algorithm which requires this knowledge.
The consensus problem can then be attacked from an engineering standpoint in an attempt to achieve a degree of reliability rather than certainty.
This reliability is comparable to the error rates in modern hardware.
However, it should scale better with the size of the system.

All of the services that we require here are provided by common network protocols such as TCP/\,IP.
Using a reliable protocol such as this can cause retransmission of some data which the algorithm does not need to proceed.
However, this is simpler than explicitly detailing all of the data for which delivery must be guaranteed and makes the algorithm easier to implement.

\subsection{Local state required}
The algorithm requires a remote object table that maps IDs to objects.
For each entry item, storage is required for the local object and the associated color.

During partial traces, each entry item is also associated with a \respo\ list, \parent, \exitl, and \depen.
The \respo\ list contains all of the remote objects from which mark gray requests are received during the first phase.
The \parent\ field contains the parent object in a tree of active objects in the first phase.
The \exitl\ contains the local exit items reachable from the entry item.
The \depen\ list contains the remote entry items from which acknowledgements are pending.

Additionally, a partial trace has storage for a root object, the \depen, \neigh, \done, \groupl, \joins, and \accepts\ lists, \mergel, and the \fset\ list.
The \depen\ list contains the partial traces from which acknowledgements are pending.
The \neigh\ list tracks the distributed objects that have been encountered by the trace but already belong to an existing trace.
The \done\ list records which partial traces have completed their first phase.

The \groupl\ list contains the known members of the current group of cooperating traces.
The \mergel\ records a potential set of group members during a group merge operation.
The \joins\ list tracks requests made by the group members to join another group; the \accepts\ list tracks requests accepted by the group members for another group to join.
The \fset\ list records which group members are on faulty machines.

\subsection{Phase one: mark gray}

The first step of the algorithm is to compute a {\em suspect subgraph}.
The suspect subgraph is the portion of the reference graph that is reachable from a suspect object, \var{x}, and contains only isolated objects.
If \var{x} is in a garbage cycle, then every remote reference to \var{x} must come from an object in the suspect subgraph.
The algorithm proceeds to the second phase only if this condition is satisfied.
The first phase likely will encounter objects that do not belong to a garbage cycle containing \var{x} because the objects reachable from \var{x} are traced rather than the objects from which \var{x} is reachable.
Locality is not violated however, since the trace of \var{x} can proceed to the next phase even when machines that do not belong to the garbage cycle fail.

Consider the case in which only one mark gray phase is active in the system at a time.
First, \var{x} computes all of the exit items that are locally reachable from \var{x} and stores this list as \exitlvar{x}.
The computation is done by the local collector during a garbage collection.
Then, \var{x} sends mark gray requests to every object contained in \exitlvar{x}.
Upon receiving a mark gray request, the procedure is replied recursively if the object is not already colored gray.

\subsubsection{Detecting termination}
\var{x} must be notified upon completion of all of the mark gray requests
generated in this phase.
This problem is commonly referred to as distributed termination detection.
Messages involved in the computation are called basic messages; messages used for termination detection are called control messages.
There exists, for any termination detection algorithm, a computation which exchanges $M$ basic messages and for which the termination detection algorithm must exchange at least $M$ control messages \cite{Tel94}.
That is, in the worst case, every termination algorithm must exchange at least as many messages as are required for the underlying computation.

Our termination algorithm is based upon the Dijkstra-Scholten algorithm for centralized computations on arbitrary networks.
We make modifications so that our computation is still correct despite an arbitrary number of system failures.
The reader should consult the original paper \cite{DS80} or Tel \cite{Tel94} for a description of the general algorithm.
We focus solely on how the termination algorithm is used in conjunction with our garbage collector.
First, we present a non-fault tolerant version of the algorithm.
Then, we show how the algorithm is modified to handle faults.

The underlying idea is to build a tree of active objects with the initiator as the root of this tree.
When an object is encountered, the object is added to the tree if it is not already present.
We do this by acknowledging every mark gray request that is sent out.
An object is colored gray when it first receives a mark gray request.
The object that sent the mark gray request is stored as the parent in the tree, but the request is not immediately acknowledged.
The object that received the mark gray request then sends out mark gray requests to all of the exit items that are locally reachable from it.
When all of these new mark gray requests have been acknowledged, an acknowledgement is sent to the parent to remove the object from the tree.
If a mark gray request is received by an object that is already gray, then an acknowledgement is immediately sent.
When \var{x} receives acknowledgements for all of its mark gray requests, the computation is terminated.

\subsubsection{Fault tolerance}
To handle faults in the algorithm, we first define what the correct outcome of the phase should be after a fault.
If a machine was faulty at the beginning of the computation, then an object reachable from that machine is colored white unless that object is reachable from \var{x} by traversing references through non-faulty machines.
This does not prevent \var{x} from proceeding to the next phase of collection unless those white objects belonged to a distributed cycle containing \var{x}.
In that case, the machine that failed must have belonged to a distributed cycle containing \var{x}.
To preserve locality, this is the only case for which a machine failure may prevent the collection of \var{x}.
Therefore, we must ensure that the same property holds even if a machine fails during the trace.
Note that if the machine containing \var{x} fails, then the trace would never have begun; we can simply mark every object white again and need not detect when this has terminated.

To ensure that an object, \var{o}, is colored white only if \var{o} is reachable from \var{x} solely by traversing through a failed machine, \var{o} maintains a list of every object from which a mark gray request was received in \respovar{o}.
If every object in this list belongs to a machine that has either failed or sent a mark white request to \var{o}, then \var{o} is colored white and sends mark white requests to every object to which \var{o} had previously sent a mark gray request.
\var{o} is removed from the computation by deleting all of the local
information stored for the trace.
Should \var{o} be waiting for acknowledgements solely from objects on machines that have failed, then \var{o} proceeds as if the acknowledgements had been received.
Unfortunately, \var{x} may incorrectly declare phase termination with this rule.
Consider an object, \var{p}, that receives mark gray requests from two different objects.
If the machine containing \var{p}'s parent, \var{q}, fails, then \var{p} is not colored white, but \var{p} can no longer prevent the termination of the phase by withholding acknowledgement from \var{q}.

This problem is solved by maintaining a separate list, \depenvar{x}, of objects from which \var{x} is waiting for acknowledgements.
Initially, \depenvar{x} is the same as \exitlvar{x}.
When some object, \var{o}, that has a parent, \var{p}, other than \var{x} receives a second mark gray request before \var{o} has acknowledged \var{p}, then \var{o} requests to be listed in \depenvar{x}.
Until \var{x} acknowledges this request, \var{o} does not send any acknowledgements.
After \var{x} acknowledges this request or \var{o} detects that the machine containing \var{x} has failed, \var{o} sets its parent to \var{x} and immediately returns acknowledgements for all of the mark gray requests, including the request from \var{p}, that \var{o} has received.
Finally, when \var{o} has received acknowledgements for all of the mark gray requests that it sends out, \var{o} returns an acknowledgement to its new parent, \var{x}, as normal.

There may be multiple paths from \var{x} to \var{o}, and a single machine failure does not necessarily break all of them.
Normally, \var{o} would withhold all but one acknowledgement to prevent \var{x} from declaring termination.
If \var{o}'s parent fails however, then \var{o} can no longer send this acknowledgement.
The only other machine that must be on all of the paths from \var{x} to \var{o} is \var{x}, so \var{o} uses \var{x} for the new parent.
This requires that \var{o} inform \var{x} that \var{x} must wait for an acknowledgement from \var{o} before declaring termination.
Since \var{o} does not send any acknowledgements before it has been added to \depenvar{x}, \var{x} cannot declare termination as long as there is at least one path to \var{o} through machines which have not failed.

Note that \var{x} may declare termination before all of the gray objects that detected their parents' failure propagate mark white requests forward.
To proceed to the next phase, \var{x} must check that all of the remote objects from which \var{x} is locally reachable are also gray.
This check can be performed during a local collection which would normally occur after the propagation.
If the local collection occurs first and the failed machines were on a garbage cycle containing \var{x}, then \var{x} incorrectly proceeds to the next phase.
We ensure that the next phase fails in this case.
Thus, the correctness of the algorithm is not compromised.
Preventing the transition to the next phase in this case is not worth the effort; to \var{x}, this case is equivalent to the case where the machines fail immediately after the beginning of the next phase.

\subsubsection{Cooperating traces}
We have been assuming that at most one mark gray phase is active at a time.
Any number of simultaneous traces can be handled by associating an ID with each trace and keeping an independent copy of the local state information associated with each trace.

Doing so incurs an extra space overhead for each object traced and leads to different mark gray phases repeatedly tracing over the same portion of the reference graph.
More efficiently, partial traces share the information from their mark gray phases.
The principle problems with this approach are forming groups of cooperating traces and detecting group termination.

The Dijkstra-Scholten algorithm generalizes to decentralized computations, computations with more than one initiator, using a technique by Shavit and Francez \cite{SF86}.
We maintain a forest of active processes instead of a tree.
A wave is passed through the root of each tree and proceeds to the next once the tree becomes empty.
When the wave is finished, termination is declared.
Any decentralized termination detection algorithm requires at least as many control messages as the number of messages, $W$, required in the best wave algorithm for that network \cite{Tel94}.
The message complexity of the Shavit-Francez algorithm is exactly $M+W$, so it is message optimal.

For a decentralized wave algorithm on an arbitrary network with $E$ channels and $N$ processes and without neighbor knowledge, the message complexity is $\Omega(E+N\log N)$ \cite{Tel94}.
In the worst case, $E$, and hence $W$ and the number of messages our algorithm uses to construct the forest if no faults occur is $O(N^2)$.
Only $O(N\log N)$ messages are used in the best case.

To cooperate, individual partial traces must agree on which partial traces belong to their group at any given time.
A partial trace rooted at \var{x} stores the list of partial traces belonging to the same group in \groupvar{\pt{x}}.
We also use \group{X} to denote \groupvar{\pt{x}}.
Each group is centralized so that groups can merge or declare termination efficiently.
A group has a designated partial trace as the leader.
The trace that \pt{x} currently considers to be the leader is stored in \leadervar{\pt{x}} and denoted \leader{x}.
If all of the traces in a group agree about the members of the group, then the leader is elected by ordering the IDs of each partial trace and choosing the smallest one.
Thus, operations which utilize the group leader may only proceed when the group is in consensus.
When using a group leader, merging groups or detecting termination operates with $O(N)$ messages.
We present each operation twice.
The first presentation does not consider faulty processes; the second presentation handles faults.

\subsubsection{Merging traces}
A partial trace, \pt{x}, is initially the only known tree in its forest.
As the trace proceeds, a list of traces that must be waited for is stored in \depenvar{\pt{x}}.
When an object, \var{o}, in \pt{x} sends a mark gray request to an object, \var{t}, that has already been colored by another trace, \pt{y}, \var{t} does
two things:
\var{o} is added to \respovar{t}, and a message \msgEC{\pt{y}} is sent to \pt{x}.
This informs \pt{x} that \pt{y} has been encountered at \var{t}.
Now, \var{t} cannot be recolored white until \pt{x} has finished all of its phases and sent a mark white request to \var{o}.
Also, \pt{x} knows that it needs to cooperate with \pt{y}.

If \pt{y} is already in \group{X}, then \group{X} is in consensus and nothing further needs to be done.
Otherwise, \pt{x} sends a merge request message, \msgMR{\leader{x}}, to \pt{y} that contains \leader{x}.
If \pt{y} responds with an \msgMR{done} message, indicating that \pt{y} has already declared termination for its mark gray phase, then \pt{x} does not need to wait for \pt{y} and does not add \pt{y} to either the \groupl\ or \depen\ list.
In this case, \pt{x} responds to \var{t}'s message with an \xmsgECACK\ message immediately.
To prevent unnecessary queries, \pt{x} keeps a list, \donevar{\pt{x}}, of encountered groups that have already declared termination.

If \pt{y} is not finished with its mark gray phase, then the two traces must come to a consensus on the new group.
First, assume that only one pair of groups attempts to merge at a time.
\pt{y} sends an accept request message, \msgAR{\leader{x}}, to notify \leader{y} that a join request is coming from \leader{x}.
When the accept request has been acknowledged with an \msgARACK{\leader{x}} message, \pt{y} then sends an acknowledgement, \msgMRACK{\leader{y}}, to \pt{x}'s merge request.
This message tells \pt{x} who \leader{y} is.
Then, \pt{x} sends a join request, \msgJR{\leader{y}}, to \leader{x} and adds \pt{y} to \depenvar{\pt{x}}.
Next, \pt{x} acknowledges \var{t}'s original \xmsgEC\ message with an \xmsgECACK\ message.

Finally, \leader{x} sends a join message, \msgJN{\group{X}}, to \leader{y}.
\leader{y} replies with an \msgAC{\group{Y}} accept message.
Each leader sets its \groupl\ field to the union of \group{X} and \group{Y} and distributes the changes to the other members of the group in \xmsgGI\ messages.
After the distribution is complete, only one trace is the new leader.
When the trace that is losing leadership of its group finishes distribution, an \xmsgMC\ message is sent to the new leader.
Once the new leader finishes distribution and receives this message, a new consensus is reached.

Since a trace does not know whether other traces in the group also wish to perform merges, we must allow multiple join or accept requests to be made of each leader.
Leaders keep a list, \joins, of join requests and a list, \accepts, of accept requests made by the group members.
When a trace is added to the group, the corresponding entry in the join or accept list is removed.
If the leader of the group changes, then the join and accept lists are forwarded to the new leader in the \xmsgMC\ message.

There are two problems with this approach.
First, if the leader processes requests in a FIFO order, then deadlock can occur.
If a cycle of traces merge with each other simultaneously, then the accept requests might all be registered before any of the join requests.
In this situation, none of the merges would proceed.
Servicing join requests before accept requests does not solve the problem; each trace might send a join request, but no one is able to reply.
The second problem occurs when two traces merge: a join request with that group may be sent to the old leader instead of the new leader.

We break the symmetry of the operations by imposing a total ordering on trace IDs to overcome the first problem.
Each trace has a unique ID generated based upon the locator of the object that started the trace.
These IDs are used in the join or accept requests to identify the trace.

Join requests are processed if the leader has no acknowledged accept requests or the join request has an ID smaller than or equal to the smallest acknowledged accept request.
There must be at least one join request that can be processed since there is always some trace with the smallest acknowledged accept request ID; another trace must hold the corresponding join request.
That other trace can have no acknowledged accept request IDs smaller than the smallest acknowledged accept request ID.
Thus, the join request ID must be smaller than or equal to the smallest acknowledged accept request ID for that trace.

The second problem is solved by forwarding \msgJN{\group{X}} messages to the new group leader when received by a trace, \pt{q}, that is not the leader.
The new leader, \leader{q}, replies to \leader{x} with an \msgAC{\group{Q}} message.
\leader{x} determines that this is a response to the original \xmsgJN\ message
because \pt{q} is a member of \group{Q}.
Note that a \xmsgJN\ message needs to be forwarded at most once, so forwarding does not affect the order of the message complexity.
The number of forwarded messages is reduced by acknowledging one accept request at a time and only acknowledging accept requests if there are no join requests.
Then, the trace that is requesting the merge does not receive permission to send a join request until there is only one join request targeting that group.
If the targeted group is also joining another, then forwarding is still required; after the targeted group has already acknowledged an accept request, a join request could be received.

\subsubsection{Fault-tolerant merging}
When a process fails, partial traces in the same group must come to a consensus about the members of the group.
If the failed process contains the group leader, then a new leader must be elected.
First, consider the case where there was no merge in progress; each partial trace's \groupl\ list is the same.
To reach a new consensus, every process in the group must ensure that it has detected the same failures as all of the other processes.
Therefore, a partial trace, \pt{x}, keeps a list of detected faults that have not been confirmed called \fsetvar{\pt{x}}.
When a fault is detected, this list is sent to \leadervar{\pt{x}}.
Once \leader{x} has received a \fset\ from each non-faulty member of the group matching \fsetvar{\leader{x}}, the leader confirms the faults to the group.

If a trace detects that the group leader has become faulty, then a new group leader is chosen from the non-faulty traces left in the group.
Messages containing the \fset\ are sent to the acting group leader.
Once all of the non-faulty group members have sent a \fset\ to the acting group leader, that trace officially becomes the new leader.

After a fault is confirmed, each trace removes the affected traces from the \groupl\ list and \fset.
The group has reached a new consensus.
If the leader changes, then all of the partial traces must reissue their join and accept requests and indicate for the accept requests whether they have already been acknowledged by the old leader.
Traces that have no pending accept or join requests send an empty list to the new leader.

Suppose that when two groups are merging, one of the group leaders becomes faulty.
Some of the traces in the groups may have different \groupl\ lists.
To avoid this problem, merging is divided into two steps.
In the first step, the new group list is distributed but not installed.
In the second step, the new group list is installed.

At the completion of the first step, we ensure that either every trace or no trace has a complete group list.
After \leader{y} sends \leader{x} the \msgAC{\group{Y}} list, both \leader{x} and \leader{y} have a copy of the complete group list.
Each leader distributes this list to their own group in \xmsgGI\ messages.
These messages are sent to group members in increasing ID order.
The next message is not sent until the previous one has been acknowledged with a \xmsgGIACK\ message.
Once a trace receives a copy of the new group list, the trace can monitor the processes of the other group for faults.
A trace, \pt{q}, stores the new group list in \mergelvar{\pt{q}} instead of \groupvar{\pt{q}} so that elections still take place from within the old
group.

Assume that a member of \group{X} faults.
The election process is executed, and a single trace collects fault reports from the other traces in \group{X} until a new leader is elected.
If the new leader has a non-empty \mergel, then distribution to the other members of the group resumes.
Otherwise, no trace in the group has a \mergel, and the merge is aborted.
\leader{y} also monitors \group{X} for faults.
Thus, when the failure of \leader{x} is detected, an \msgMI{\leader{y}} message is sent to the acting leader of \group{X}.
The acting leader responds with an \msgMIACK{merging} message if \leader{y} is in its \mergel.
Otherwise, the acting leader responds with \msgMIACK{canceled}.
If \leader{y} receives \msgMIACK{canceled}, then the merge is aborted.
The situation is symmetric for a fault in \group{Y}.

Once \leader{x} distributes the new group list to all of the group members, a \xmsgCR\ message is sent to \leader{y}.
After \leader{y} distributes the new group list to all group members, a \xmsgCRACK\ message is sent in reply.
The second step begins now.
Each leader sends \xmsgCM\ messages to the group members in descending order of ID.
Again, the leader waits for a message to be acknowledged with a \xmsgCMACK\ message before sending the next one.
When a trace receives a \xmsgCM\ message, \groupl\ is replaced with \mergel.
As before, the old leader sends an \xmsgMC\ message containing the join and accept requests to the new leader.

Suppose that \leader{X} faults during this distribution.
Only some of the group members have the new group list stored in their \groupl\ field.
If those members elect a new leader from \group{Y}, then the acting leader of \group{X} does not receive a {\fset} from each member.
The acting leader has a non-empty \mergel\ however, so \leader{y} is contacted to have those messages forwarded.
When all of the forwarded messages are received, \xmsgCM\ messages are sent to the traces still in \group{X}.
If the new leader is in \group{X}, then each trace sends with the \fset\ a flag indicating whether \mergel\ is empty.
The acting leader sends \xmsgCM\ messages to those with a non-empty \mergel.

If every member of \group{X} had already received a \xmsgCM\ message and the new leader would be in \group{Y}, then \leader{y} does not confirm the faults until there are no more \xmsgCM\ messages to distribute.
If the new leader is in \group{X} however, then the new acting leader in \group{X} does not receive fault sets from traces in \group{Y} that have
not received a \xmsgCM\ message.
Each trace in \group{Y} records the faults detected in \group{X} prior to receiving a \xmsgCM\ message.
When a \xmsgCM\ message is received, the recorded faults are added to \fset\ and sent to the acting leader.
Thus, the acting leader of \group{X} eventually receives a \fset\ from everyone in both groups.
Again, the situation is symmetric for a fault in \group{Y}.

\subsubsection{Group termination detection}
When a group is formed, a token, \term, is created for termination detection.
The token is initially held by the leader and is passed when all of the mark gray requests made by the trace are acknowledged.
The token contains two lists of partial traces: \termvar{done} and \termvar{next}.
When all of the mark gray requests for a trace are acknowledged, the trace is added to the \termvar{done} list, \depen\ is added to the \termvar{next} list, and the token is forwarded to the first trace in the \next\ list.
If the \next\ list is empty, then the token is sent to the leader of the group.
The leader sends the \termvar{done} list to all of the group members if the group is in consensus.
Each group member adds the traces in \termvar{done} to the \done\ list and removes them from the \depen\ list.
When their \depen\ list is empty, the traces proceed to the next phase.

A trace that is dependent on a failed trace cannot tell when all of the mark gray requests sent by the failed trace have been handled.
Instead, the trace keeps a list of all of the objects in neighboring traces to which mark gray requests have been sent.
For example, \pt{x} would store in \neigha{\pt{x}}{\pt{y}} a list of the objects that sent \pt{x} an \msgEC{\pt{y}} message.
If \pt{x} detects that the machine \pt{y} is on has failed, \pt{x} sends a re-mark gray request, \msgRG{\pt{x}}{\pt{y}}, to each object in \neigha{\pt{x}}{\pt{y}}.
When an object first receives a re-mark gray request, a re-mark gray request is sent to each object in \exitl, and the trace that sent the re-mark gray request adds the object to \depen.
An object that has already received a re-mark gray request reports the new trace to which it belongs, \pt{z}, and is added to \neigha{\pt{x}}{\pt{z}}.
This procedure retraces portions of the reference graph only when the initiator of a trace fails.
Objects that have not detected the failure of \pt{y} buffer re-mark gray requests until the failure is confirmed.

Traces that have declared that all of their mark gray requests are acknowledged may send out additional mark requests if a fault is detected.
Thus, when a fault is detected for a trace in the \donevar{term} or \nextvar{term} lists, the token is discarded, and the leader of the group
creates a new token.
The leader detects this condition by having each trace send whether it has received the token when the \fset\ is reported.

\subsection{Phase two: mark black}

The second phase is based upon a back tracing algorithm \cite{ML97} but is restricted to the suspect subgraph that was computed in the first phase.
Additional modifications are made to support cooperation between traces.

After the first phase, the initiator of a trace, \var{x}, examines \entryl\ to see if a mark gray request was received for each incoming reference.
If not, then \var{x} is reachable from outside of the suspect subgraph and is not collected.
The algorithm proceeds directly to the third phase in that case.

If several traces have cooperated to compute their suspect subgraphs, then \var{x} might receive mark gray requests from another trace, \pt{y}, even though \var{y} is not reachable from \var{x}.
If \var{y} was reachable, then \pt{x} would have encountered \pt{y} in the first phase, and \pt{y} would be in \depenvar{\pt{x}}.
Since \pt{y} sent \var{x} a mark gray request, \pt{x} is in \depenvar{\pt{y}}.
Therefore, neither trace would have declared termination of the first phase until they were both done.
Thus, both traces must have been in the \done\ list of the same token.
The \done\ list of the token that declared \pt{x} terminated provides a conservative estimate of the partial traces in \var{x}'s suspect subgraph.
The list is stored in \marksvar{\pt{x}}, and all of the objects that were colored gray by one of these traces is the {\em conservative suspect subgraph}.

A better determination of the suspect subgraph is made by storing the \depen\ list of each partial trace in the token.
The strongly connected components of the graph this induces on the partial tracings is then computed.
This is still not sufficient to determine whether a mark gray request came from an object reachable from \var{x}; \pt{x} can meet \pt{y} past the location of \var{y}.
There need not be a path from the meeting point of the two traces to any object, including \var{y}, encountered by \pt{y} before the meeting point.
Relying on the conservative estimate makes the cooperation of traces in the second phase more complicated than if an exact determination is known for objects that either reach or are reachable from \var{x}.

\subsubsection{The basic trace}
We first describe the behavior of the algorithm in the case that at most one trace is in the second phase at a time.
If \var{x} receives mark gray requests from all of its incoming references and each mark request was sent by a partial trace in the \marks\ list, then \var{x} is colored black, and a mark black request containing the \marks\
list is sent to each exit item of \var{x}.

An exit item, \ex{q}, that receives a mark black request computes all of the entry items from which it is locally reachable.
This can be done during the next local collection \cite{ML97} and is stored in \backlvar{\ex{q}}.
If any entry item in \backl\ is white, then the exit item is \state{live}.
Thus, \backl\ is essentially the same as an \exitl\ computed in the first phase with the additional knowledge of whether an exit item is reachable from a white entry item.

If \ex{q} is not reachable from a white entry item, then each entry item in \backl\ that was colored gray by a trace in the \marks\ list but not colored
black is colored black and sends a mark black request to the corresponding exit item.
Each mark black request returns whether the exit item is \state{live} or \state{garbage}.
If any entry item receives a reply of \state{live}, then \ex{q} returns \state{live} to the object from which the mark black request was received.
Otherwise, \ex{q} returns \state{garbage}.
\var{x} is \state{garbage} if every exit item returns a reply of \state{garbage}.
In that case, the algorithm proceeds to the third phase and reclaims all of the objects that were colored black in the second phase.

Either every object receiving a mark black request belongs to a distributed garbage cycle or no such cycle exists.
Therefore, preserving the property of locality does not require that we ensure that the algorithm continues to propagate mark black requests when a machine fails.
Termination is detected by waiting for every mark black request to be acknowledged.
Fault tolerance is achieved by assuming that processes that fail or receive a mark white request are \state{live}.
A process can receive a mark white request if, as described in the first phase, a process containing a gray object fails.

\subsubsection{Cooperating traces}
We maintain a group of cooperating back traces using the same fault-tolerant protocol for maintaining a group of cooperating forward traces in the previous phase.
The initial group is the \marks\ list.
The \neigh\ list are the objects that were colored black by other traces in the same group.
Groups do not merge.
An entry item that has incoming references that sent mark gray requests for a trace outside of the \marks\ list safely block the current trace.
The trace waits for the reference to either send a mark white request or be deleted.
The trace returns \state{live} in the first case; the algorithm continues in the second case.
Similarly, if an exit item is locally reachable from an entry item, \ei{o}, that was colored either gray or black by a trace outside of the \marks\ list, then the current trace blocks until either \ei{o} is colored white or \var{o} is collected.
Again, the exit item returns \state{live} in the first case; the algorithm continues in the second case.

As with the mark gray phase, termination of the mark black phase for a group is detected by passing a token.
An additional piece of information is needed: whether the objects that were colored black are actually garbage.
Unlike the basic trace, an initiator does not determine by itself if the objects that are colored black are garbage even if all of the mark black requests return \state{garbage}.
Each trace stores in the token whether an object reachable from outside of the conservative suspect subgraph was encountered.
If every back trace completes without discovering an object reachable from outside of the subgraph, then every object colored black is garbage.
Conversely, if every back trace discovers an object that is reachable from outside of the suspect subgraph, then every object colored black was live when the trace began.
If some of the back traces encountered live objects however, then whether the objects colored black are garbage is unclear.
The back trace is not confined to a single strongly connected component.

To resolve this ambiguity, the initiator of a back trace computes whether a particular object encountered in the back trace is reachable from an object outside of the back trace.
Thus, the \entryl\ of each black entry item and the \backl\ of each exit item that received a mark black request is returned in addition to a reply of \state{live} or \state{garbage}.
These lists form a subgraph of the reference graph and are replaced by any graph with the same transitive closure to reduce network message size.
The smallest such graph is called the {\em transitive reduction} and is computed as efficiently as the transitive closure \cite{AGU72}.

After the mark black requests are handled, suppose that some of the traces did not encounter an object reachable from outside of the conservative suspect subgraph.
The reachability information returned by the back trace is used to compute whether each of these traces should return \state{live} or \state{garbage}.

A trace, \pt{x}, sends \neigha{\pt{x}}{\pt{y}} to each \pt{y} that was encountered by \pt{x}.
The \pt{y}s examine their reachability information and respond with the objects in \neigha{\pt{x}}{\pt{y}} that are known to be reachable from an object outside of the conservative suspect subgraph.
Meanwhile, \pt{x} receives \neigh\ lists from other traces.
If the response from a neighboring trace indicates that an additional neighbor in these lists is \state{live}, then \pt{x} forwards the additions to the lists returned earlier.
Termination detection is done by the standard Dijkstra-Scholten algorithm.

This process is similar to retracing the portion of the conservative suspect subgraph from which \var{x} is reachable.
The number of messages is significantly reduced: the computation is locally performed by the initiator of each of the cooperating traces; the other objects in the trace are not involved.
In the worst case, if $N$ is the number of cooperating initiators, then two messages must be sent for each neighbor and $O(N^2)$ messages are required to distribute the initial neighbor lists.

Finally, if a fault occurs in one of the initiators, then the remaining traces must ensure that they can still determine reachability from the portion of the reference graph for which the failed initiator's back trace was responsible.
This is done by querying the neighbors directly instead of retracing since the neighbors already have the reachability information for their children.
Each neighbor is sent a re-mark black request.
A neighbor is attached to the first trace from which it receives such a request.
Then, those neighbors that that neighbor and its children have encountered are added to the attached trace's main neighbor lists.
If more than one trace has failed, then the process is repeated as necessary.
Should the retracing stage have begun already, then retracing is restarted once all of the re-mark black requests are handled.
Neighbors that were already determined to be \state{live} are still considered \state{live}.

\subsection{Reference modification barrier}
We must detect reference graph modifications by the mutator to prevent partial traces from succeeding incorrectly.
Reference deletions are not tracked: these can never add a root from which the suspect subgraph is reachable.
However, reference creations must be tracked.
Modifications do not need to be detected during the mark gray phase, since only objects reachable from the initiator are encountered, but they must be detected during the mark black phase.

We use the color purple to indicate objects that may have had references added.
When a remote reference is added to an object, the object is colored purple.
Local references are expensive to track in this manner.
Since the trace succeeded, the object must have been isolated.
Thus, the only way to add a local reference to the object is to traverse to an object in the suspect subgraph of the same machine.
Therefore, if a remote reference to a gray or black object is traversed, then the object is colored purple.

If a trace encounters a purple object during phase two, then the trace returns that the object is \state{live}.

\subsection{Reference transfer barrier}
Ideally, subsequent traces reuse the \exitl\ computed for an object.
To ensure that \exitl\ is correct, we determine if no exit item in the exit list could have been locally reachable since the last computation of \exitl.
When an exit item is added to the exit list, that exit item must be isolated.
Thus, an exit item was locally reachable only if some object from which the exit item was reachable was passed as a parameter or as a return value in a remote procedure call.
We use a transfer barrier to trap such calls\cite{ML97}.

When a distributed object, \var{x}, gains a local root from a remote procedure call, \exitlvar{x} is examined, and if there is an object, \var{y}, such that \exitlvar{x} and \exitlvar{y} intersect, then \exitlvar{y} is marked as out-of-date and \exitlvar{x} is discarded.
The \var{y}s are enumerated using the inverse image of each {\exitl}; these contain the same information and require only a constant factor more space.
If \var{y} is isolated the next time that a trace passes through it, then \exitl\ is recomputed.

Mark gray requests that arrive during the recomputation wait before propagating additional mark gray requests.
However, synchronization with other processes is not required.
Similarly, a mark black request waits for the computation to complete before propagating additional mark black requests.
If the transfer barrier is applied to an object that is computing a new \exitl, then the barrier is first applied to any existing \exitl.
The application of the transfer barrier is recorded so that the barrier is reapplied when the new \exitl\ is available.
Thus, only the \exitl\ of the object that the barrier is applied to is discarded instead of all of those that were marked as out-of-date.

\section{Proposed heuristic}

Developing a good heuristic is vital for reducing the amount of unnecessary work the distributed garbage collector performs.
A heuristic compatible with our design goals satisfies the following five properties:
\begin{enumerate}
\item
For the garbage collector to be complete, the heuristic must eventually suspect all garbage objects even in the presence of arbitrarily contrived system faults and local mutator operations.
\item
For the garbage collector to be prompt, the heuristic should suspect an object soon after it becomes garbage.
Preferably, this should be determined using only locally available information rather then requiring message passing.
\item
For the garbage collector to have a high degree of locality and only a small amount of remote synchronization, the heuristic should rely only on remote information that is already required for the tracing process.
The heuristic should aggressively avoid performing work on or suspecting objects that are live.
\item
For the garbage collector to make good use of non-scalable resources, the heuristic should not impose significant tasks on the local collector or use significantly more space than the tracing process requires.
\item
For the garbage collector to have little synchronization with the mutators, the heuristic should not impose significantly more overhead then the local collector requires for basic reference operations: creation, assignment, dereferencing, and deletion.
\end{enumerate}

\subsection{Examination of existing heuristics}

We now reconsider the naive and distance heuristics presented earlier and show that neither meets the requirements listed above.

\subsubsection{Naive heuristic}

The naive heuristic suspects a significant number of live objects which requires a large amount of remote communication.
A live object generates an arbitrarily large number of messages even if the object is never referenced.
This is unacceptable for systems with limited bandwidth between nodes or for systems with large numbers of objects.
Introducing delays between re-suspecting the same object reduces the number of messages by at most a constant factor.
Additionally, the delay in reclaiming garbage increases proportionally with the reduction in messages; the collection process is no longer prompt.
A generational system further reduces the number of unnecessary messages.
However, it has an even larger decrease in promptness of collection.
It is clear that a system based upon the naive heuristic can generate an unbounded number of messages even if no object in the system is actually garbage or is being used by a mutator.

\subsubsection{Distance heuristic}

The distance heuristic improves upon the naive heuristic by placing an upper bound upon the number of messages passed by live objects that are not being used by a mutator.
If there are no changes to the reference graph, then the distance heuristic eventually reaches a steady-state condition in which no messages are passed.
However, the distance heuristic has a number of drawbacks and fails to meet our requirements.

The distance heuristic can fail to be prompt.
Once an item becomes unreachable, the heuristic requires a number of messages and local collections on the order of the threshold parameter before that item becomes suspect.
The distance heuristic requires a great deal of unnecessary remote communication.
Messages are passed to all reachable systems even if many of those machines are not required to identify a suspect item.
The distance heuristic performs a great deal of work on live objects.
A single modification to the reference graph may result in as many messages being transmitted as the number of objects multiplied by the threshold parameter.

\subsection{Introduction of new heuristic}

We create an entirely new heuristic to work with our tracing scheme.
This heuristic works in strong cooperation with the tracer to infer as much information as possible and avoids unnecessary work duplication or message passing.
Although there may be many objects on a garbage cycle, only one needs to initiate a trace for all of them to be reclaimed.
To be complete, our heuristic ensures that at least one object on every potential garbage cycle can become suspect.
The other members of that potential garbage cycle then rely on this object to initiate collection and avoid becoming suspect.
We adapt this idea to maintain completeness even in the presence of system faults and local mutator actions.
This reduces message passing prior to tracing and reduces the amount of work being performed on live objects.

Our desired heuristic properties transform into a set of goals for the new heuristic.

\begin{itemize}
\item
The heuristic must have at least one member of every garbage cycle that can become suspect.
\item
The heuristic should quickly suspect at least one member of an unrooted garbage cycle using minimal message passing.
\item
The heuristic should avoid initiating communication with another machine.
\item
The heuristic should use only a constant factor more time than local collection and a constant factor more space than tracing.
\item
The heuristic should impose no barriers on mutator reference manipulation.
\end{itemize}

An unfortunate side effect is that by reducing the number of messages passed, the heuristic becomes more complicated by the need to infer the state of the world from only local information and information used by the tracer.
The need for safety in the presence of system faults and local mutator operations complicates the individual operations.
Therefore, it is likely that the heuristic can be further optimized with respect to the number of live objects suspected without requiring that additional information be computed.

\subsection{Local state required}

The heuristic requires two bits of storage in each entry item \ei{x}.
One bit, \vulnevar{\ei{x}}, has states {\em vulnerable} and {\em invulnerable}.
\ei{x} should be vulnerable if \ei{x} may need to become suspect to initiate collection.
\ei{x} should be invulnerable if another entry item in the potential garbage cycle can reliably initiate collection.
The other bit, \activvar{\ei{x}}, has states {\em active} and {\em inactive}.
\ei{x} should be active if every exit item from which \ei{x} is reachable may have been traced.
An entry item should be inactive if at least one exit item from which \ei{x} is reachable has not been traced.

The heuristic also requires a parallel array, \entrycvar{\ei{x}}, to the entry list.
The entries of \entrycvar{\ei{x}} consist of a single color associated with each item in the entry list.

\subsubsection{Additional local requirements}

The heuristic requires the following information already being computed elsewhere: the current color of the entry item, the entry list of an entry item, the exit list of an entry item, the ability to be notified of changes in network connectivity to a remote machine that is currently connected, the ability to determine an object's reachability from a local root.
We initially develop heuristics that are simple but incomplete.
These heuristics may not require all of this information.

\subsection{Conditions for trace initiation}

An object is said to be {\em eligible} if it is vulnerable, active, and isolated.
Eligible objects are considered suspect if they are not currently being traced.
Suspect objects should initiate a trace without undue delay for the garbage collection to be prompt.
These properties are determined locally without significant computation.

\subsection{Simple Heuristic}

The first heuristic we develop is referred to as the Simple Heuristic.
Although it is not suitable for all garbage collection tasks, this heuristic is later extended to support general collection.
Garbage is collected under the assumption that no faults occur and that the actions of the local mutators are restricted.

The first local mutator restriction is that exit items reachable from a colored entry item on the same local machine must be isolated.
A function call that passes through a colored entry item causes that entry item to temporarily have a local root.
However, every exit item on the same local machine reachable from that entry item is then locally reachable.
Thus, our restriction implies that it is not possible for a function call to be invoked on a colored entry item.
The second local mutator restriction is that no new remote references that point to a colored entry item can be created.

The following observation is used repeatedly in the justification of the heuristic.
Suppose that we have an initial entry item, \ei{x}, from which a trace starts.
For the heuristic to consider the traced entry items again, they must not have been reclaimed by the tracer.
Since we are assuming that no faults occur, it therefore was impossible to reclaim those items using information available at that time.

\subsubsection{Determination of entrycolors}

When a trace reaches an isolated entry item, \ei{o}, that has not been colored, \ei{o} is colored gray.
Also, we initialize \entrycvar{\ei{o}} to be the colors of \var{o}'s entry list members.
Note that if \ei{o} is already gray, it is currently involved in some trace, and \entrycvar{\ei{o}} has already been initialized.

\ei{o} is no longer part of a trace when the collector marks it white.
From the time that \entrycvar{\ei{o}} is initialized until the time that \ei{o} is colored white, the darkest color reached by each entry list member
of \ei{o} is recorded.
Due to the assumptions that we have made, no entry item can be colored purple.
Therefore, we only consider the following ordering on colors from darkest to lightest: black, gray, and white.
This color information determines the state transitions made by \var{o}.

\subsubsection{First termination case}

We consider first the case that every item in \entrycvar{\ei{o}} is non-white at termination.
Since \ei{o} was not reclaimed, \ei{o} must be reachable from an exit item that was not traced.
Let \ex{q} be one such exit item and \ei{q} be the entry item to which \ex{q} refers.
Every item in \entrycvar{\ei{o}} is non-white, so \ei{q} cannot be \ei{o}.
From just this information, we cannot determine if \ei{q} has been traced.
It therefore is possible that \ei{q} is still live, and the reference graph can be modified so that \ei{o} is no longer reachable from \ex{q}.
Thus, we do not know whether \ei{o} is still reachable from an exit item that has not yet been traced.
By the state definitions we have made, \ei{o} is active.

The next consideration is whether \ei{o} is vulnerable.
To show that \ei{o} is invulnerable, we demonstrate that there is an entry item that can reliably initiate collection of \var{o}.
It cannot be the case that every entry item from which \ei{o} is reachable has every item in its \entryc\ colored.
Were this to occur, we would have definite knowledge that \var{o} was garbage.
By the assumptions we have made, \var{o} would have been reclaimed, and the heuristic would not be able to again consider \ei{o}.
Thus, there exists an entry item \ei{m} such that \ei{o} is reachable from \ei{m}, \entrycvar{\ei{m}} has at least one uncolored item, and every
entry item on the path from \ei{m} to \ei{o} has no uncolored items in its \entryc.

We do not permit function calls to pass through colored entry items.
Thus, once an entry item has been colored, it can never become locally reachable.
We require also that every exit item reachable from a colored entry item on the same machine be isolated.
Therefore, once an entry item is colored, it is not possible to change the exit items that are locally reachable from that entry item.
Furthermore, the local mutator may not break the path between the entry item and one of these exit items.
This implies that if there is some path between two entry items for which every entry item traversed was colored, then that path can never be broken except through collection.

Thus, there is no way to break all of the paths from \ei{m} to \ei{o} except through collection.
We therefore rely on \ei{m} to initiate collection for \var{o}.
Suppose that there was a circular dependency such that \var{m} was relying on \ei{o} to initiate collection.
\var{m} relies on another entry item only if \ei{m} was traced and every
item in \entrycvar{\ei{m}} was colored.
However, if \ei{m} was traced, then it is not possible for new references to be made to \ei{m}.
It is also not possible to uncolor the items in \entrycvar{\ei{m}}.
Thus, every item in \entrycvar{\ei{m}} must still be colored.
If this is the case, then \var{o} cannot rely on \ei{m}, a contradiction.
Therefore, there is no such circular dependency.

\subsubsection{Second termination case}

Next, consider the case that at least one item in \entrycvar{\ei{o}} is white.
We know immediately from this information that there is an exit item, \ex{q}, that has not been traced and from which \ei{o} is reachable.
For \ex{q} to have been deleted, it would have had to notify \ei{o}, which would have removed it from the entry list.
However, \ex{q} must still be in the entry list for it to have an entry in \entrycvar{\ei{o}}.
Thus, \ei{o} is reachable from an untraced exit item.
By the state definitions we have made, \ei{o} is inactive.

We now construct a situation in which no entry item other than \ei{o} can reliably initiate collection to show that \ei{o} is vulnerable.
Let \var{q}, \var{o}, and \var{x} be objects with remote references from \var{q} to \var{o}, from \var{o} to \var{x}, and from \var{x} to \var{o}.
Suppose that \var{o} and \var{x} are isolated and have no additional remote references.
Additionally, suppose that \var{q} has a local root.

If we start a trace at either \ei{o} or \ei{x}, then both \ei{o} and \ei{x} are colored gray.
It is not possible to traverse \ei{q} going forward from \ei{o} or \ei{x}.
Thus, \ei{q} was not traced.
Since \var{q} has a local root, the reference from \var{q} to \var{o} can be deleted leaving \var{o} and \var{x} on an unrooted cycle.
From the first termination case, \ei{x} is invulnerable after this is done.
Therefore, no entry item besides \ei{o} can reliably initiate a collection for \var{o}.
Hence, \ei{o} is vulnerable to preserve completeness.

\subsubsection{Handling of inactive entry items}

Let \ei{o} be an inactive entry item.
From the two termination cases, \ei{o} must have been traced and must have at least one item in its \entryc colored white.
\ei{o} was traced, so our local mutator assumptions prevent the creation of new remote references that point to \var{o}.
As long as one item in \entrycvar{\ei{o}} is uncolored, \ei{o} remains inactive by the argument presented above.
Suppose that every item in \entrycvar{\ei{o}} is either colored or reclaimed.
At this point, we do not know if there are any untraced exit items from which \ei{o} is reachable.
The example given in the previous section contains no untraced exit items from which \ei{o} is reachable.
Thus, \ei{o} is active to preserve completeness.
Whether \ei{o} is vulnerable is determined as before.

\subsection{Fault Heuristic}

The second heuristic we develop is referred to as the Fault Heuristic.
This heuristic is an extension of the Simple Heuristic although it still does not support general garbage collection.
The primary difference between the Simple Heuristic and the Fault Heuristic comes from removing the assumption in the Simple Heuristic that no faults occur.
We retain the two given restrictions for local mutator actions: exit items reachable from a colored entry item must be isolated and no new remote references that point to a colored entry item are created.

\subsubsection{Supporting fault-tolerant tracing}

The first set of changes we make to the Simple Heuristic support fault-tolerant tracing.
In fault-tolerant tracing, it is possible for a trace to prematurely abort due to a fault.
Suppose that we have an initial entry item \ei{x} from which a trace starts.
If no faults occur during the trace, then we can proceed exactly as specified in the Simple Heuristic.
Otherwise, assume that at least one fault has occurred while tracing.

We show first that \ei{x} cannot be invulnerable or inactive.
Suppose that \var{x} is on a garbage cycle and \ei{x} is the only entry item in the garbage cycle that is able to initiate a collection.
If \ei{x} starts a trace that encounters a failure, it is still true that no other entry item in the garbage cycle can initiate a trace.
Additionally, if \ei{x} is either inactive or invulnerable, it is not able to start a future trace.
The entry items in the garbage cycle are in a steady-state condition and cannot be changed without the intervention of the heuristic.
Therefore, at least one entry item in the garbage cycle needs to be made active and vulnerable to ensure that the garbage cycle is collected.

The only entry item that is guaranteed to know the outcome of the trace is \ei{x}.
Consider what is required if \ei{x} is not active and vulnerable.
It is possible for an arbitrary number of machines involved in the trace to encounter failure with each machine knowing only about its own fault.
To conservatively protect our ability to collect the garbage cycle, we are forced to give the ability to start a future collection to each of the machines that encountered a fault.
In order to limit the number of machines that can start a new trace in response to the faults, we have to engage in some message passing between the machines.
Neither alternative is desirable.

There is no way for \var{x} to reasonably be collected by another object without additional message passing by either the tracer or the heuristic.
Therefore, we leave \ei{x} vulnerable and active to retry the trace again.
Better strategies exist, but we expect faults to be an atypical occurrence and are willing to accept reduced performance in this case for a simpler heuristic.
One strategy would be to implement a message passing system to notify \ei{x} when the fault was likely to have subsided.

We show now that all of the entry items encountered in the trace started by \ei{x}, except for \ei{x}, can be treated exactly as in the Simple Heuristic.
Suppose that \ei{n} was an entry item reached in the trace from \ei{x} and that \var{n} and \var{x} are distinct.
As demonstrated in the Simple Heuristic, there is a path from \ei{x} to \ei{n} containing only colored entry items; this path cannot be broken through local mutator actions.
Therefore, no matter what state \ei{n} was left in, the system is consistent once \ei{x} successfully completes its trace.
There is no guarantee that \ei{n} receives notification that the trace started by \ei{x} has faulted.
Thus, we cannot restrict the state of \ei{n} in the Fault Heuristic more than we can restrict the state in the Simple Heuristic.

\subsubsection{Supporting other kinds of faults}

The second set of changes we make to the Simple Heuristic support completeness in the presence of machine faults.
Suppose that a machine containing \ex{x}, \machine{x}, experiences a fault.
At some point in the future, the machine containing \ei{x}, \machine{i}, is guaranteed to receive notification of this fault.
How we handle this fault notification depends on whether the connection between \ex{x} and \ei{x} was persistent.

Suppose that the connection between \ex{x} and \ei{x} was persistent.
Then, if \machine{x} ever reestablishes the connection with \machine{i}, \machine{x} must have preserved all of the object graph from which \ei{x} was
reachable.
When the connection between \machine{x} and \machine{i} is down, no trace can propagate across that link.
Therefore, if \ei{x} was invulnerable, there is still some object from which \ei{x} is reachable that can reliably initiate a collection.
If \ei{x} was inactive, then the status of \ex{x} can not change while the connection between \machine{x} and \machine{i} is down.
We treat this case exactly as in the Simple Heuristic.
Therefore, we take no action in response to the connection between \machine{x} and \machine{i} being terminated if the connection was persistent.

Next, suppose that the connection between \ex{x} and \ei{x} was non-persistent.
If \machine{x} ever reestablishes the connection with \machine{i}, then \machine{x} must have preserved none of the object graph from which \ei{x} was reachable.
Therefore, the status of \ei{x} cannot be changed by the reestablishment of the connection between \machine{i} and \machine{x}.
When the connection between \machine{x} and \machine{i} goes down, we know immediately that \ex{x} will be deleted.
This presents a difficulty: \ei{x} may have been relying on an entry item, \ei{m}, to initiate collection for which the only path between \ei{m} and \ei{x} passed through \ex{x}.
\ei{x} is therefore vulnerable to conservatively maintain completeness.
Whether \activvar{\ei{x}} changes as a result of \ex{x} being deleted is handled exactly as in the Simple Heuristic.

\subsection{Complete Heuristic}

The final heuristic we develop is referred to as the Complete Heuristic.
We no longer maintain the two restrictions on local mutator actions.
We allow exit items reachable from a colored entry item to be locally reachable.
We also allow new remote references to be created that point to a colored entry item.
If neither of these conditions occur, then we have already shown how to handle collection in the Simple and Fault Heuristics.

The Complete Heuristic uses state information about exit items in determining when to start traces.
We do not require exit items to store information, but that approach may lead to easier implementation.
The idea used repeatedly in this heuristic is that a trace passes through an exit item only if that exit item is isolated.
Therefore, we require the ability to initiate traces on a local machine when an exit item on that machine transitions from being locally reachable to being isolated.
This ability is used to lift our local mutator restrictions.

\subsubsection{Trace continuations}

Suppose that a trace encounters an entry item \ei{x}.
The trace has been successfully propagated when the collector determines every isolated exit item locally reachable from \ei{x} and passes a notification of the trace to those exit items.
However, if there is a locally reachable exit item reachable from \ei{x}, then there are situations in which our current heuristics fail.

Let \var{o} be an object on a different machine than \var{x} with remote references from \var{x} to \var{o} and from \var{o} to \var{x}.
Suppose that \var{o}, \var{x}, and \ex{x} are isolated.
If \ei{o} and \ei{x} start traces while \ex{o} is locally reachable, then \ei{o} is inactive; \ei{x} is invulnerable.
No entry item in the garbage cycle is able to start a collection.
Since all of the objects in the cycle are garbage, there is no point in the future at which we are able to collect them.

However, if \ex{o} is no longer locally reachable, then a trace started by \ei{x} or \ei{o} is sufficient to collect the garbage cycle.
In essence, we wish to continue one of the traces that would have passed through \ex{o}.
Unfortunately, it is impractical to maintain the arbitrarily large history of traces to support future continuations; our space requirements are violated.
Instead, we start a new trace if, from the information available to the heuristic, it may be necessary to do so to collect a garbage cycle.

\subsubsection{Restoration of eligibility}

We detail now an algorithm to restore eligibility to selected entry items in response to a locally reachable exit item becoming isolated.
Together with the other developed heuristics, this is sufficient for general garbage collection.
Suppose that \ex{q} is a locally reachable exit item that becomes isolated.

First, we trace backwards from \ex{q} to find all of the entry items from which \ex{q} is locally reachable.
During this back trace, we consider only isolated objects.
If we encounter a live object during the back trace, then \ex{q} is also live.
Therefore, no trace can pass through \ex{q}, and we must wait until \ex{q} becomes isolated again.
Denote the list of entry items from which \ex{q} is locally reachable as \precuvar{q}.
The remainder of this heuristic is the consideration of cases.

\subsubsection{First precursor case}

Suppose that some entry item, \ei{a}, in \precuvar{q} is eligible.
Then, \ei{a} is able to start a trace at some point in the future.
Additionally, there is no advantage to having a trace go through \ex{q} until \ei{a} initiates its trace.
In order for the second phase of tracing to succeed, then \ei{a} must first be traced.

We know that unless some local mutator action has broken the connection between \ei{a} and \ex{q}, that a trace going through \ei{a} proceeds through \ex{q}.
If the connection between \ei{a} and \ex{q} is broken, then there must be a modification of the reference graph between \ei{a} and \ex{q}.
However, this implies that \ex{q} is locally reachable.
When \ex{q} becomes isolated again, we are guaranteed to have another application of this heuristic to choose a new entry item from \precuvar{q} to rely upon.
Thus, actions taken by the local mutator do not affect our ability to collect garbage.

\subsubsection{Second precursor case}

Now, suppose that no entry item in \precuvar{q} is eligible.
Since we only consider entry items that are isolated, every entry item in \precuvar{q} is either inactive or invulnerable.
As before, if any modification of the reference subgraph between an entry item in \precuvar{q} and \ex{q} occurs, then \ex{q} is locally reachable.
This guarantees that the heuristic is reapplied when \ex{q} becomes isolated again.
Thus, local mutator actions cannot violate the properties of correctness, completeness, or fault-tolerance if those properties are preserved in the case where no local mutator actions occur.
Therefore, without loss of generality, we assume that no local mutator actions occur during this procedure.

The \exitl\ of each entry item in \precuvar{q} is scanned.
If \ex{q} is not present in at least one \exitl, then one of the entry items, \ei{b}, is chosen.
We reset the eligibility of \ei{b} by making it active and vulnerable.
By the first precursor case, we are now finished.

\subsubsection{Third precursor case}

Otherwise, every entry item in \precuvar{q} contains \ex{q} in its \exitl.
While \ex{q} is locally reachable, it is possible that the only entry item that can start a trace of the garbage cycle containing \var{q} is removed from the cycle.
Suppose that \var{m} and \var{n} are objects on the same local machine and that \var{o} is an object on another machine.
Also, let there be remote references from \var{m} and \var{n} to \var{o} and from \var{o} to \var{m} and \var{n}.
Suppose that the only object able to start a collection is \var{m} and that \ex{o} is locally reachable.
While \ex{o} is locally reachable, it is possible that all of the paths from \var{m} to \ex{o} can be broken by the local mutator.
Then, the only entry item, \ei{n}, from which \ex{o} is reachable contains \ex{o} in its \exitl.

We know that no entry item in the garbage cycle can initiate a collection, so some entry item in the garbage cycle must become eligible to preserve completeness.
This allows us to rely on objects contained on other machines for collection without worrying that the reference graph between those two machines has changed.
One way to avoid this situation is to maintain the value of \precuvar{q} that was computed when \ex{q} last was isolated.
Denote this list as \oldprvar{q}.
If \precuvar{q} and \oldprvar{q} are identical, we know that no entry item was removed while \ex{q} was locally reachable.
Therefore, the situation that was outlined above cannot occur.

\subsubsection{Fourth precursor case}

Finally, we have the case that \oldprvar{q} and \precuvar{q} are different.
We have no information stored in \ex{q} to indicate which entry items have traced through it.
Additionally, there is no guarantee that the entry items that traced through \ex{q} have any record of that trace at an arbitrary point in the future.
Therefore, the only alternative is to make eligible some entry item from which \ex{q} is locally reachable.
However, this is exactly as occurred in the second precursor case.

\section{Future Work}
The next step is to implement this system and see how it performs in an actual long-lived distributed system.
We plan to implement our algorithm using Java in cooperation with the existing uniprocessor collector.
We will test our algorithm against several of the cyclic distributed garbage collectors described in this paper.
This will give a greater assurance that the algorithm is correct and may provide insights for further enhancements.

It would also be useful to compare the algorithm with existing cyclic distributed garbage collection algorithms in terms of performance.
We know of no systematic performance evaluations of existing algorithms
\cite{Rod98} despite the dependency many have upon a heuristic.
This is likely because cyclic distributed garbage collection is not yet a common feature in distributed systems.
Testing is hindered because real applications are the only meaningful sources of memory usage patterns.
Assumptions about the random distribution of memory accesses are likely gross generalizations, since real programs do not behave randomly \cite{Wil95}.

We expect to see several improvements using our system.
The message overhead for active portions of the graph is likely to be smaller than with the distance heuristic, which sends a barrage of messages throughout the system whenever local roots are added or removed.
Our heuristic allows local roots to be added to and removed from an object without message passing if the reachability between entry and exit items is not changed.
Furthermore, our system is likely to be more prompt for smaller cycles of garbage, such as those created by simple client-server relationships, since only a small number of messages would be exchanged before the objects became suspect.
We also expect that the number of messages sent for our traces will be smaller than in back tracing, since our traces can cooperate to avoid repeated work.
It is unknown however, if overlap is common enough to warrant the overhead that this requires.

\section{Acknowledgements}

Special thanks to Dr. James D. Arthur for providing guidance during a
 substantial portion of the development and for encouraging us to continue when
 things weren't working.
Equipment support provided by the Virginia Polytechnic Institute and State
 University Systems Research Center.
Thanks to everyone who provided feedback and suggestions for improving this
paper.

\section{Definitions}

\define{Active}{
A condition possessed by an entry item when all of the exit items from which the entry item is reachable might already have been traced.}
\define{Complete}{
A garbage collector that always reclaims garbage.}
\define{Correct}{
A garbage collector that reclaims only garbage.}
\define{Distributed object}{
An object to which remote references point.
These objects have exactly one entry item associated with them, and each remote reference is from an exit item.}
\define{Eligible}{
An entry item that can start a trace.}
\define{Entry item}{
An item containing the entry list for a distributed object and any state information for traces through that object.}
\define{Entry list}{
The list of processes that contain a remote reference to a particular object.}
\define{Exit item}{
The encapsulation of a remote reference to an object.}
\define{Fault-tolerant}{
The ability to maintain an invariant despite the occurrence of specific incidents denoted as faults.}
\define{Garbage}{
An object that is not reachable from any root.}
\define{Garbage collection}{
The identification and removal of garbage from a system.}
\define{Host}{
The machine containing the physical storage for a particular distributed object.}
\define{Inactive}{
A condition possessed by an entry item when at least one exit item from which the entry item is reachable has not been traced.}
\define{Invulnerable}{
A condition possessed by an entry item when another entry item in the potential garbage cycle can reliably initiate collection.}
\define{Isolated}{
An object that is not locally reachable but still live.}
\define{Local object}{
An object to which there are no remote references.}
\define{Local root set}{
All objects in the root set belonging to a particular machine.}
\define{Locality}{
A measure of the number of machines whose cooperation is required to collect a garbage object in relation to the number of machines on that garbage cycle.}
\define{Locally reachable}{
An object for which there exists a path from a member of a particular set to the object that does not traverse any remote references.
If no specific set is mentioned, then the set of local roots is assumed.}
\define{Mutator}{
A process independent of the garbage collector that can create or delete references.}
\define{Non-scalable resource}{
A resource that has a super-linear cost function.}
\define{Object reference graph}{
A directed graph with vertices corresponding to the objects in the system and edges corresponding to inter-object references.}
\define{Pseudo-root set}{
The set of entry items present on a particular machine.
Although the elements of the pseudo-root set are not actually roots, they are treated as roots by the local collector.}
\define{Reachable}{
An object for which there exists a path from a member of a particular set to the object.
If no specific set is mentioned, then the root set is assumed.}
\define{Root set}{
A subset of the vertices of an object reference graph that have no incoming edges and are specially designated by the system.}
\define{Suspect}{
An entry item that is believed to be garbage and should be traced.}
\define{Suspect subgraph}{
The portion of the reference graph that is reachable from a suspect object and contains only isolated objects.}
\define{Vulnerable}{
A condition possessed by an entry item when the entry item might need to become suspect to initiate collection.}

\bibliography{dgcnotes}

\end{document}